\documentclass[12pt]{article}
\usepackage{amsmath}
\usepackage{graphicx,psfrag,epsf}
\usepackage{enumerate}
\usepackage[longnamesfirst]{natbib}
\usepackage{url} % not crucial - just used below for the URL
\usepackage{latexsym,bm}
\usepackage{amsthm}
\usepackage{booktabs}
\usepackage{epsfig}
\usepackage{graphicx}
\usepackage{amsfonts}
\usepackage{amssymb}
\usepackage{multirow}
\usepackage{algorithm}
\usepackage{comment}
\usepackage{color}
\allowdisplaybreaks
\newcommand{\blind}{0}

\begin{document}
\theoremstyle{definition} \theoremstyle{corol}
\theoremstyle{proposition} \theoremstyle{condition}
\newtheorem{theorem}{Theorem}
\newtheorem{example}{Example}
\newtheorem{remark}{Remark}
\newtheorem{lemma}{Lemma}
\newtheorem{definition}{Definition}
\newtheorem{corol}{Corollary}
\newtheorem{condition}{Condition}
\newtheorem{proposition}{Proposition}
\newtheorem{ass}{Assumption}

\def\spacingset#1{\renewcommand{\baselinestretch}%
{#1}\small\normalsize} \spacingset{1}

%%%%%%%%%%%%%%%%%%%%%%%%%%%%%%%%%%%%%%%%%%%%%%%%%%%%%%%%%%%%%%%%%%%%%%%%%%%%%%

\if0\blind
{
  \title{\bf Adaptive Testing for Alphas in High-dimensional Factor Pricing Models}
  \author{Qiang Xia
    \hspace{.2cm}\\
   College of Mathematics and Informatics, South China Agricultural University\\
    and \\
    Xianyang Zhang\thanks{ Address correspondence to Xianyang Zhang (zhangxiany@stat.tamu.edu)} \\
    Department of Statistics, Texas A\&M University}
  \maketitle
} \fi

\if1\blind
{
  \bigskip
  \bigskip
  \bigskip
  \begin{center}
    {\LARGE\bf Title}
\end{center}
  \medskip
} \fi

\bigskip
\begin{abstract}
This paper proposes a new procedure to validate the multi-factor pricing theory by testing the presence of alpha in linear factor pricing models with a large number of assets. Because the market's inefficient pricing is likely to occur to a small fraction of exceptional assets, we develop a testing procedure that is particularly powerful against sparse signals. Based on the high-dimensional Gaussian approximation theory, we propose a simulation-based approach to approximate the limiting null distribution of the test. Our numerical studies show that the new procedure can deliver a reasonable size and achieve substantial power improvement compared to the existing tests under sparse alternatives, and especially for weak signals.
\end{abstract}

\noindent%
\vfill

\newpage
\spacingset{1.8} % DON'T change the spacing!
\section{Introduction}
\label{sec:intro}

As one of the most fundamental results in empirical finance, the factor pricing model postulates how financial returns are related to market risks and has many critical applications such as portfolio selection, fund performance evaluation, and corporate budgeting. There are several well-known factor models, including the capital asset pricing model (CAPM) proposed by \cite{Sharpe1964} and \cite{Lintner1965}, the arbitrage pricing theory (APT) model by \cite{Ross1976}, the Fama-French three-factor model by \cite{Fama1993}, and the Fama-French five-factor model by \cite{Fama2015}.

Let $y_{it}$ be the excess return of the $i$th asset at time $t$ and $X_t=(x_{1t},\dots,x_{rt})^\top\in\mathbb{R}^{r\times 1}$ be a vector of observable factors such as tradable market risk factors. The form of the factor pricing model is given by
\begin{align}\label{Test1}
y_{it}=\alpha_i+\beta^\top_i X_t+u_{it},\quad i=1,2,\dots,N,\quad t=1,2,\dots,T,
\end{align}
where $\beta_i\in \mathbb{R}^{r\times1}$ is a vector of factor loadings and $u_{it}$ represents the idiosyncratic error which is assumed to be independent of the factors $X_{t}$ with $\mathbb{E}[u_{it}]=0$ and $\text{cov}(u_{it},u_{jt})=\sigma_{ij}$ for $i,j=1,\ldots, N$.
The multi-factor pricing theory implies that the vector of intercepts $\boldsymbol{\alpha}=(\alpha_1,\dots,\alpha_N)^\top$ should be zero. Therefore, to empirically validate the factor pricing theory, researchers often consider the problem of testing
\begin{align*}
H_0: \boldsymbol{\alpha}=\mathbf{0} \quad \text{versus} \quad H_a: \boldsymbol{\alpha}\neq \mathbf{0}.
\end{align*}
An example of a commonly used test for this problem is the exact multivariate $F$-test proposed by \cite{Gibbons1989}, which assumes Gaussian errors and $N<T$. However, its application has been confined to a relatively small number of portfolios using monthly returns observed over a relatively long period. With the advent of modern financial markets, where thousands of securities are traded daily, there is a need for tests that can handle a large or diverging number of assets $N$. Consequently, recent efforts have focused on developing tests that allow $N$ to be larger than the time dimension $T$, given the popularity of high-dimensional factor models. Some notable attempts along this direction include \cite{Pesaran2012}, \cite{Gungor2013}, \cite{Pesaran2017}, \cite{Ma2020}, and \cite{Giglio2021}.

The above methods are all sum-of-squares types, generally having good power performance against dense alternatives. For instance, \cite{Pesaran2012,Pesaran2017} proposed Wald-type tests for validating the capital asset pricing model based on the diagonal elements of the sample covariance matrix. However, according to \cite{Fan2015}, inefficient market pricing is more likely to occur in exceptional assets rather than systematic mispricing of the entire market. Therefore, it is desirable to develop a test that has good power against sparse alternatives.
The Wald-type test can suffer from low power when $\boldsymbol{\alpha}$ is a sparse vector (i.e., sparse alternatives). To address this issue, \cite{Fan2015} proposed adding a power enhancement component to the Wald test (referred to as the FLY test) to improve its power under sparse alternatives. The power enhancement component approaches zero under the null hypothesis (so it does not affect the null distribution of the test asymptotically)
but diverges and dominates the Wald statistic under some specific regions of sparse alternatives.
%Power enhancement of the FLY test for sparse alternatives is achieved when $\alpha_i/\sigma^{1/2}{ii} > 3 \log{\log(T)}\sqrt{\log(N)/T}$.
Another effective way to deal with sparse alternatives is the maximum type test proposed by \cite{Gungor2016}, although they did not provide an in-depth analysis of the test. Recently, \cite{Feng2021} re-examined the maximum type statistic and proved its asymptotic null distribution to be the type I extreme value distribution.

In general, the maximum type statistic only considers the largest entry of $\boldsymbol{\alpha}$ and is not adaptive to the unknown sparsity level. This paper proposes a novel testing procedure that overcomes the limitations of existing methods and achieves considerable power improvement. The new test is designed to
\begin{enumerate}
    \item be adaptive to the unknown sparsity level;
    \item borrow the cross-sectional dependence information to boost the statistical power.
\end{enumerate}
These features enable the test to outperform existing solutions over a wide range of settings. To motivate our approach, we derive a likelihood ratio test (LRT) under the assumption of joint normality for the errors against a sparse alternative that assumes precisely $k$ nonzero components in $\boldsymbol{\alpha}$. The resulting test statistic captures both the sparsity level of $\boldsymbol{\alpha}$ and the cross-sectional dependence via the error precision matrix. Unfortunately, computing the test statistic involves solving a set optimization problem with complexity $O(N^k)$, which is computationally infeasible for moderate $k$. To mitigate this issue, we propose replacing the error precision matrix with a diagonal matrix in the test statistic, which reduces computational complexity to $O(N\log N)$ for any fixed $k$.

An important consideration in this framework is how to select the appropriate number of non-zero components $k$ that are adaptive to the signals. We propose choosing $k$ to maximize the standardized test statistic, which draws inspiration from the literature on adaptive testing (\cite{Ingster1993}; \cite{Fan1996}; \cite{Donoho2004}). To implement the testing procedure, we develop a simulation approach to approximate the empirical null distributions of the test statistics, based on recent high-dimensional Gaussian approximation theory.

For a high-dimensional Central Limit Theorem (CLT) and bootstrap theory for independent and identically distributed (i.i.d) sequences, we refer readers to \cite{Chernozhukov2017}, while for a CLT for high-dimensional time series, we refer readers to \cite{Chang2021} (also see \cite{Zhang2018}; \cite{ZhangW2018}). In addition to the problem considered in this paper, Gaussian approximation theory has been applied to other large-scale testing problems, such as inference for high-dimensional mean (\cite{Chang2017a}), high-dimensional covariance matrices (\cite{Chang2017b}), high-dimensional precision matrices (\cite{Chang2018}), high-dimensional white noise testing (\cite{Chang2017c}), and inference for high-dimensional linear regression (\cite{ZhangC2017}; \cite{Dezeure2017}).

We demonstrate theoretically that our simulation approach accurately approximates the null distribution and that the proposed test is consistent against a class of sparse alternatives. While our procedure is motivated by the LR test derived under the joint normality assumption of the errors, the validity of our approach does not depend on the Gaussian assumption. Our empirical results show that the proposed procedure maintains reasonable size and achieves considerable power improvement compared to existing tests under weak and sparse alternatives. Furthermore, unlike other tests, the number of factors in our proposed test can diverge as $N$ and $T$ increase.

The paper is organized as follows. Section 2 introduces the test statistic and its adaptive version, and proposes a simulation-based procedure to approximate the limiting null distributions. Section 3 investigates the asymptotic properties of the tests. Sections 4 and 5 are dedicated to simulation studies and real data analysis, respectively. Technical arguments are presented in the Appendix.

Finally we introduce some notation that will be used throughout the paper. For $a\in\mathbb{R}$, let $\lfloor a\rfloor$ be the integer part of $a$.
For a vector $z=(z_1,\ldots,z_N)\in\mathbb{R}^N$ and $q>0$, define $|z|_q =(\sum_{i=1}^N|z_i|^q)^{1/q}$ with $|z|=|z|_2$. Set $|z|_0=\sum^{N}_{i=1}\mathbf{1}\{z_i\neq 0\}$ which is the number of nonzero elements in $z$, and $|z|_\infty =\max_{1\leq i\leq N}|z_i|$. For $V=(v_{ij})\in \mathbb{R}^{N\times N}$, define the matrix norms: $\|V\|_1=\max_{1\leq j\leq N}\sum_{i=1}^N|v_{ij}|$, $\|V\|_2=\max_{|z|=1}|Vz|$, $\|V\|_\infty=\max_{1\leq i\leq N}\sum_{j=1}^N|v_{ij}|$ and $|V|_\infty=\max_{1\leq i,j\leq N}|v_{ij}|$. Write $\text{diag}(V)=\text{diag}(v_{11},\ldots,v_{NN})$. Let $\|\cdot\|_0$ be the cardinality of a set. For $S\subseteq\{1,\ldots,N\}$, let $V_{S,S}$ be the submatrix of $V$ that contains the rows and columns in $S$. Similarly, define $V_{S,-S}$ as the submatrix of $V$ with the rows in $S$ and the columns in $\{1,\ldots,N\}\setminus S$.

\section{Methodology}\label{sec:meth}
\subsection{Test statistics}
Let $Y_t=(y_{1t},\ldots,y_{Nt})^\top$ and $\boldsymbol{\beta}=(\beta_1,\ldots,\beta_N)^\top$. Then the factor pricing model (\ref{Test1}) can be written in the matrix form as
\begin{align}\label{Test3}
Y_t=\boldsymbol{\alpha}+\boldsymbol{\beta} X_t+U_t,\quad t=1,2,\dots,T,
\end{align}
where $U_t=(u_{1t},\ldots,u_{Nt})^\top$ is a martingale difference sequence satisfying $\mathbb{E}(U_t|\mathfrak{F}_{t-1})=0$
with $\mathfrak{F}_s$ denoting the $\sigma$-field generated by $\{U_t\}_{t\leq s}$.
Note that $\{\boldsymbol{\alpha}\in \mathbb{R}^N:\boldsymbol{\alpha}\neq \mathbf{0}\}=\cup_{k=1}^N \mathcal{A}_{k},$ where $\mathcal{A}_{k}=\{\boldsymbol{\alpha}\in \mathbb{R}^N:|\boldsymbol{\alpha}|_0=k\}$.
To motivate the subsequent derivations, we consider the testing problem:
\begin{align*}
H_0:\boldsymbol{\alpha}=\mathbf{0}\quad \text{versus} \quad H_{a,k}: \boldsymbol{\alpha}\in \mathcal{A}_{k}
\end{align*}
where the sparsity level $k$ under the alternative is assumed to be known as a priori.
In Section \ref{sec:k}, we shall introduce a data-driven procedure to select $k$.

Under the Gaussian assumption $U_t\sim \mathbb{N}(\mathbf{0},\Sigma)$, the negative log-likelihood (up to a constant term) is equal to
$$\mathcal{L}(\boldsymbol{\alpha};\boldsymbol{\beta};\Gamma)=\frac{1}{2}\sum_{t=1}^T(Y_t-\boldsymbol{\alpha}-\boldsymbol{\beta}X_t)^\top \Gamma(Y_t-\boldsymbol{\alpha}-\boldsymbol{\beta}X_t),$$
where $\Gamma=\Sigma^{-1}=(\gamma_{i,j})$ is the precision matrix. 
Fixing $\Gamma$ and using Lemma~2 in the appendix, we derive the likelihood ratio test (LRT) for testing $H_0$ against $H_{a,k}$,
which takes the form of
\begin{align*}
T\underset{\|S\|_0=k}{\max}\{\Gamma \widehat{\boldsymbol{\alpha}}\}^\top_S\Gamma_{S,S}^{-1}\{\Gamma \widehat{\boldsymbol{\alpha}}\}_S.
\end{align*}

Define
$\mathbf{X}=(X_1,\ldots,X_T)\in\mathbb{R}^{r\times T}$,
$\mathbf{Y}=(Y_1,\ldots,Y_T)\in\mathbb{R}^{N\times T}$ and $\mathbf{1}=(1,\ldots,1)^\top\in\mathbb{R}^{T\times 1}$, where $\bar{Y}=T^{-1}\sum_{t=1}^TY_t$ and $\bar{X}=T^{-1}\sum_{t=1}^TX_t.$
Note that the MLE of $\boldsymbol{\alpha}$ can be expressed as
$$\widehat{\boldsymbol{\alpha}}=(\widehat{\alpha}_1,\ldots,\widehat{\alpha}_N)^\top=\bar{Y}-\widehat{\boldsymbol{\beta}}\bar{X},$$
where $\widehat{\alpha}_i=T^{-1}\sum_{t=1}^Ty_{it}\{1-(X_t-\bar{X})^\top\mathbf{W}\}$ with $\mathbf{W}=\{T^{-1}\sum_{t=1}^T(X_t-\bar{X})(X_t-\bar{X})^\top\}^{-1}\bar{X}\in\mathbb{R}^{r\times 1}$.
Here $\widehat{\boldsymbol{\beta}}$ is the MLE of $\boldsymbol{\beta}$ given by $\widehat{\boldsymbol{\beta}}=(\mathbf{Y}- \bar{Y}\mathbf{1}^\top)(\mathbf{X}- \bar{X}\mathbf{1}^\top)^\top\left\{(\mathbf{X}-\bar{X}\mathbf{1}^\top)(\mathbf{X}- \bar{X}\mathbf{1}^\top)^\top\right\}^{-1}.$
Under the independence between $U_t$ and $X_t$, it can be shown that the variance of $\widehat{\alpha}_i$
converges in probability to
$\sigma^2_i=T^{-1}(1+\mathbb{E}X_t^\top\Sigma_X^{-1}\mathbb{E}X_t)\sigma_{ii},$
which can be estimated consistently by
$\widehat{\sigma}^2_i=T^{-2}\big(1+\bar{X}^\top\mathbf{W}\big)\sum_{t=1}^T\widehat{u}^2_{it},$
where $\widehat{u}_{it}=y_{it}-\widehat{\alpha}_i-\widehat{\boldsymbol{\beta}}^\top_iX_t$.

Next, we introduce a procedure to estimate the unknown precision matrix. Following \cite{Fan2015}, we define a screening set:
$$\hat{\textbf{s}}=\{i:|\widehat{\alpha}_i|>\widehat{\sigma}_i\delta_{N,T}, i=1,\ldots,N\},$$
where $\delta_{N,T}=C\log(\log T)\sqrt{\log N}$ with $C$ being some positive constant. Under a minimal signal assumption,
we can prove the sure screening property (see Lemma 5 in the appendix) that
$\hat{\textbf{s}}$ contains all the nonzero components of $\boldsymbol{\alpha}$ with probability converging to one. Set $$\widetilde{\boldsymbol{\alpha}}=\widehat{\boldsymbol{\alpha}}I(\hat{\textbf{s}})=\Big(\widehat{\alpha}_1I(|\widehat{\alpha}_1|>\widehat{\sigma}_1\delta_{N,T}),
\ldots,\widehat{\alpha}_NI(|\widehat{\alpha}_N|>\widehat{\sigma}_N\delta_{N,T})\Big)^\top$$
and let $\widehat{U} = \mathbf{Y} - \widetilde{\boldsymbol{\alpha}}\mathbf{1}^\top-\widehat{\boldsymbol{\beta}}\mathbf{X}$ and $\widehat{\Sigma}_U=\widehat{U}\widehat{U}^\top/T=(\widehat{\sigma}_{uij})^{N}_{i,j=1}$.
Further write $V=\textmd{diag}^{1/2}(\Sigma)$ and $R$ as the correlation matrix of $\Sigma$. Accordingly, we define
$\widehat{V}=\textmd{diag}^{1/2}(\widehat{\Sigma}_U)$ and $\widehat{R}$ as the sample correlation matrix.
Following \cite{Chen2013} and \cite{Shu2019},
we use a slightly modified version of the graphical lasso estimator:
\begin{align}\label{est0}
&&\widehat{\textmd{K}}_\rho=\arg\min_{\Psi>0}\{\textmd{tr}(\Psi\widehat{R})-\log\det(\Psi)+\rho|\Psi^-|_1\},
\end{align}
where the minimization is over the set of positive-definite matrices $\Psi=(\psi_{i,j})$, $\rho$ is a non-negative tuning parameter, and $|\Psi^-|_1=\sum_{1\leq i\neq j\leq N} |\psi_{i,j}|$.
Then we estimate $\Gamma$ by $\widehat{\Gamma}=\widehat{V}^{-1}\widehat{\textmd{K}}_\rho\widehat{V}^{-1}$.

\begin{remark}{\rm
To estimate the sparse precision matrix, we can also use the nodewise Lasso [\cite{Meinshausen2006}] or the Constrained $l_1$-Minimization
for Inverse Matrix Estimation (CLIME) [\cite{CLL2011}].}
\end{remark}

Combining the above derivations, we obtain the test statistic in the form of
\begin{align}\label{integer}
G_T(k)=T\underset{\|S\|_0=k}{\max}
\{\widehat{\Gamma}\widehat{\boldsymbol{\alpha}}\}^\top_S\widehat{\Gamma}_{S,S}^{-1}\{\widehat{\Gamma}\widehat{\boldsymbol{\alpha}}\}_S.
\end{align}
However, computing the value of $G_T(k)$ is challenging as it involves a set optimization problem with computational complexity of order $O(N^k)$. This complexity makes computation prohibitive even for moderate values of $k$. In the following, we propose an approach to address this issue.

\subsection{Modified tests}\label{sec:mod}
The computational cost of (\ref{integer}) is high, especially when combined with the simulation-based approach in Section \ref{sec:sim}. To reduce the computational burden, we propose a modified test by replacing $\widehat{\Gamma}_{S,S}$ with $\text{diag}(\widehat{\Gamma}_{S,S})$, which includes only the diagonal elements of $\widehat{\Gamma}$. This eliminates the need for best subset selection, significantly improving computational efficiency. The resulting test statistic is given by:
\begin{align}\label{Mtest}
\widetilde{G}_{T}(k)=T\underset{\|S\|_0=k}{\max}\{\widehat{\Gamma} \widehat{\boldsymbol{\alpha}}\}^\top_S\text{diag}^{-1}(\widehat{\Gamma}_{S,S})\{\widehat{\Gamma} \widehat{\boldsymbol{\alpha}}\}_S.
\end{align}
Let $\widehat{\Gamma} \widehat{\boldsymbol{\alpha}}=(\widehat{z}_i)^{N}_{i=1}$ and $\widehat{\Gamma}=(\widehat{\gamma}_{i,j})_{i,j=1}^N$. Sort the values $|\widehat{z}_j|^2/\widehat{\gamma}_{j,j}$
in descending order, say $|\widehat{z}_{j_1^*}|^2/\widehat{\gamma}_{j_1^*,j_1^*}\geq |\widehat{z}_{j_2^*}|^2/\widehat{\gamma}_{j_2^*,j_2^*}\geq \cdots \geq |\widehat{z}_{j_N^*}|^2/\widehat{\gamma}_{j_N^*,j_N^*}$. Direct calculation yields that
\begin{align*}
\widetilde{G}_{T}(k)=T\sum_{i=1}^k\frac{\widehat{z}_{j_i^*}^2}{\widehat{\gamma}_{j_i^*,j_i^*}}.
\end{align*}
Therefore, the maximization in (\ref{Mtest}) can be solved using sorting algorithms, which leads to the time complexity of the order $O(N\log N)$ for any fixed $k$.

\subsection{Selection of $k$ and adaptive testing}\label{sec:k}
 In this section, we propose a data-dependent approach to select $k$, inspired by the literature on adaptive testing. For clarity, we focus on $\widetilde{G}_{T}(k)$ but emphasize that the same idea applies equally to $G_T(k)$. Let $\widehat{\mathbb{E}}[\widetilde{G}_{T}(k)]$ and $\widehat{\text{var}}[\widetilde{G}_{T}(k)]$ be the estimates of the mean and variance of $\widetilde{G}_{T}(k)$ under the null, which can be constructed using the simulation-based approach in Section \ref{sec:sim} below. Our idea is to find $k$ which maximizes the standardized statistic, i.e.,
\begin{align*}
k_0=\arg\max_{1\leq k\leq K}\frac{\widetilde{G}_T(k)-\widehat{\mathbb{E}}[\widetilde{G}_{T}(k)]}{\sqrt{\widehat{\text{var}}[\widetilde{G}_{T}(k)]}},
\end{align*}
where $K$ is a user-specified upper bound. The corresponding adaptive test statistic is given by
\begin{align*}
\widetilde{\mathcal{G}}_T(K)=\frac{\widetilde{G}_T(k_0)-\widehat{\mathbb{E}}[\widetilde{G}_{T}(k_0)]}{\sqrt{\widehat{\text{var}}[\widetilde{G}_{T}(k_0)]}}=\max_{1\leq k\leq K}\frac{\widetilde{G}_T(k)-\widehat{\mathbb{E}}[\widetilde{G}_{T}(k)]}{\sqrt{\widehat{\text{var}}[\widetilde{G}_{T}(k)]}}.
\end{align*}

\subsection{A simulation-based approach}\label{sec:sim}
To approximate the sampling distributions
of $G_T(k)$ and $\widetilde{G}_{T}(k)$ for $k \geq1$ and $\widetilde{\mathcal{G}}_T(K)$ under sparsity assumption, we propose a simulation-based approach, which is closely related to the multiplier bootstrap in \cite{Chernozhukov2017}, \cite{Belloni2018} and \cite{Chang2021}.
A key step in deriving the simulation-based procedure is to understand the estimation effect of
$\widehat{\boldsymbol{\beta}}$ in $\widehat{\boldsymbol{\alpha}}$. We note that
\begin{align*}
\widehat{\boldsymbol{\alpha}}=\bar{Y}-\widehat{\boldsymbol{\beta}}\bar{X}
=&\,\boldsymbol{\alpha}-T^{-1}\big(\mathbf{U}- \bar{U}\mathbf{1}^\top\big)\big(\mathbf{X}- \bar{X}\mathbf{1}^\top\big)^\top\mathbf{W}+\bar{U}\\
 \approx &\,\boldsymbol{\alpha}+T^{-1}\sum^{T}_{t=1}U_t[1-(X_t-\bar{X})^\top\Sigma_X^{-1}\mathbb{E}X_t]:=\boldsymbol{\alpha}+\widetilde{Z},
\end{align*}
where $\mathbf{U}=(U_1,\ldots,U_T)$.
Under the independence between $\mathbf{U}$ and $\mathbf{X}$, it can be shown that the covariance matrix of $\sqrt{T}\widetilde{Z}$ converges in probability to
$(1+\zeta)\Sigma$ for $\zeta=\mathbb{E}X_t^\top\Sigma_X^{-1}\mathbb{E}X_t$. The core of the simulation approach is to construct a quantity using simulated normal variables that approximates the distribution of $\widetilde{Z}$.
The details of our procedure can be described as follow:
 \begin{itemize}
\item [\rm 1.]
Generate a sequence of independent standard normal random variables $\{\varepsilon_t\}^{T}_{t=1}$ that are independent of the sample, and let $$\widetilde{Z}^*=T^{-1}\sqrt{1+\bar{X}^\top\widehat{\Sigma}_X^{-1}\bar{X}}\sum_{t=1}^T\left\{\big(Y_t-\bar{Y}\big)
-\widehat{\boldsymbol{\beta}}\big(X_t-\bar{X}\big)\right\}\varepsilon_t,$$
where $\widehat{\Sigma}_X=T^{-1}\sum_{t=1}^T(X_t-\bar{X})(X_t-\bar{X})^\top$.
\item [\rm 2.]
Compute the simulation-based statistics for %$G_T(k)$ and
$\widetilde{G}_{T}(k)$ as
\begin{eqnarray*}
&&\widetilde{G}^*_{T}(k)=T\underset{\|S\|_0=k}{\max}\{\widehat{\Gamma} \widetilde{Z}^*\}^\top_S\text{diag}^{-1}(\widehat{\Gamma} _{S,S})\{\widehat{\Gamma}  \widetilde{Z}^*\}_S.
\end{eqnarray*}
\item [\rm 3.]
Repeat Steps 1-2 $B$ times to get the $1-\alpha$ quantiles of %$\{G^*_{T,j}(k)\}^{B}_{j=1}$ and
$\{\widetilde{G}^*_{T,j}(k)\}^{B}_{j=1}$,
which are the simulation-based critical values of %$G_T(k)$ and
$\widetilde{G}_{T}(k)$.
\item [\rm $3'$.]
We estimate the mean and variance of $\widetilde{G}_T(k)$ by
\begin{align*}
&\widehat{\mathbb{E}}[\widetilde{G}_{T}(k)]=B^{-1}\sum^{B}_{j=1}\widetilde{G}_{T,j}^*(k),\\
&\widehat{\text{var}}[\widetilde{G}_{T}(k)]=B^{-1}\sum^{B}_{j=1}\left\{\widetilde{G}_{T,j}^*(k)-B^{-1}\sum^{B}_{j=1}\widetilde{G}_{T,j}^*(k)\right\}^2,
\end{align*}
respectively. To approximate the distribution of the adaptive test, we consider
\begin{align*}
\widetilde{\mathcal{G}}_{T,j}^*(K)=\max_{1\leq k\leq K}\frac{\widetilde{G}_{T,j}^*(k)-\widehat{\mathbb{E}}[\widetilde{G}_{T}(k)]}{\sqrt{\widehat{\text{var}}[\widetilde{G}_{T}(k)]}}.
\end{align*}
The $1-\alpha$ quantiles of $\{\widetilde{\mathcal{G}}^*_{T,j}(K)\}^{B}_{j=1}$ will be the critical value of our adaptive test.
\end{itemize}

\section{Theory}
\subsection{Consistency under the null}\label{sec:null}
In this subsection, we study the theoretical properties of the proposed tests and justify the
validity of the simulation-based approach. Define $\alpha(T)=\sup_{A\in\mathcal{F}^0_{-\infty},B\in\mathcal{F}_T^{\infty}}|P(A)P(B)-P(AB)|,$
where $\mathcal{F}^0_{-\infty}$ and $\mathcal{F}_T^{\infty}$ denote the $\sigma$-algebras generated by $\{(X_t, U_t):  t \leq 0\}$ and
$\{(X_t, U_t): T\leq t \}$ respectively. Let $m=\underset{1\leq i\leq N}{\max}\|\{\gamma_{ij}, j\neq i, 1\leq j\leq N\}\|_0$.
For a random variable $\xi$, we define its Orlicz norm to be
$\|\xi\|_{\psi_{\gamma_1}} = \inf\{q\geq 0:\mathbb{E}[\psi_{\gamma_1}(|\xi|/q)]\leq1\}$ with $\psi_{\gamma_1}(x):=\exp(x^{\gamma_1})-1$.
To facilitate the derivations, we make the following assumptions. Below we let $c_i$ with $0\leq i\leq 6$ be some positive constants independent of $N$ and $T$.

\begin{ass}\label{assum-1}{\rm~
\begin{itemize}
\item [\rm (a)] Suppose $\{U_t\}^{T}_{t=1}$ is a sequence of martingale difference sequence with $\mathbb{E}(U_t)=0$ and $\text{cov}(U_t)=\Sigma$, where $\lambda_{\min}(\Sigma)\geq c_1>0$. Assume that there exists a sequence of constants $B_T > 1$ and a
universal constant $\gamma_1 \geq1$ such that $\|u_{it}\|_{\psi_{\gamma_1}}\leq B_T$ for all $i$ and $t$.
   % $\sup_{b\in \mathbb{V}^{N-1}}\|b^\top U_t\|_\iota < c_2$ and
%$\sup_{b\in \mathbb{V}^{N-1}}\|b^\top \Gamma U_t\|_\iota < c_3$ for some
%constants $c_2, c_3> 0$, where $\mathbb{V}^{N-1}=\{b \in \mathbb{R}^{N\times 1}: |b| = 1\}$.
Moreover, assume that  $\sup_{1\leq i,j\leq N}\text{var}(u_{it}u_{jt})<\infty$, $\|\Sigma\|_1<\infty$ and $\|\Gamma\|_1<\infty$.
\item [\rm (b)] Suppose $\{(X_t, U_t)\}^{T}_{t=1}$ is $\alpha$-mixing with the mixing coefficient $\alpha(t)$.
There exist $c_2, c_3>0$ such that for all $t\in \mathbb{Z}^+$,
$$\alpha(t)\leq \exp(-c_2t^{c_3}).$$
\item [\rm (c)] Suppose $\{X_t\}^{T}_{t=1}$ is strictly stationary and is independent of $\{U_t\}^{T}_{t=1}$.
There exist $\omega>0$ and $c_4>0$ with $3c^{-1}_4+c^{-1}_3>1$ such that
$$P(|x_{it}|>s)\leq\exp\left\{-(s/w)^{c_4}\right\}.$$
\item [\rm (d)] Suppose $\lambda_{\min}(\Sigma_X)\geq c_5>0$
and $\underset{1\leq i\leq N}{\max}\|\beta_i\|<c_6$ for some $c_5,c_6>0$.
\end{itemize}
}
\end{ass}

\begin{remark}{\rm In contrast to the i.i.d assumption on $\{U_t\}^{T}_{t=1}$ in Assumption 4.1 (i) of \cite{Fan2015} and (A1) of \cite{Feng2021}, we relax this assumption in our paper because we utilize the high-dimensional CLT from \cite{Chang2021} [see Theorem~7 therein].
Assumption~\ref{assum-1}(a) implies that $\mathbb{E}\{\exp(|u_{it}|^{\gamma_1}B_T^{-\gamma_1})\}\leq2$,
which is equivalent to the condition that the tail of $u_{it}$ satisfy $P(|u_{it}|>u)\leq2\exp\left\{-u^{\gamma_1}B_T^{-\gamma_1}\right\}.$}
Assumptions 1 (b) (d) are identical to those in \cite{Fan2015}. Specifically, conditions (b) and (c) require the tail decay and strong mixing assumptions, respectively, which allow us to use Bernstein-type inequalities for weakly dependent sequences in our technical proofs.
\end{remark}
Denote $\mathcal{C}_{N,T,k}=\max\big\{B^2_Tk^{\frac{2+6c_3}{c_3}}(\log N)^{\frac{1+2c_3}{c_3}}, B^3_Tk^{10}(\log N)^{\frac{7}{2}},k^{\frac{6-2c_3}{c_3}}(\log N)^{\frac{3-c_3}{c_3}}\big\}$.
\begin{theorem}\label{theo1} Assume $\mathcal{C}_{N,T,k}= o(T^{1/3})$ and $k^2(r^{3}+r^2m)(\log NT)^{5/2}/\sqrt{T}=o(1)$. Then under Assumption \ref{assum-1}~(a)-(d) and the null hypothesis $H_0$, we have
\begin{align}\label{Test8}
\underset{\upsilon\geq0}{\sup}\bigg| P\Big(\widetilde{G}^*_T(k)\leq\upsilon\Big|X_1^T,Y_1^T\Big)-P\Big(\widetilde{G}_T(k)\leq\upsilon\Big)\bigg|=o_p(1),
\end{align}
where $X_1^T=\{X_1,\ldots,X_T\}$ and $Y_1^T=\{Y_1,\ldots,Y_T\}$.
\end{theorem}
\begin{theorem}\label{theo2} Assume $\mathcal{C}_{N,T,k}= o(T^{1/3})$ and $K^4(r^{3}+r^2m)(\log NT)^{5/2}/\sqrt{T}=o(1)$. Under Assumption \ref{assum-1}~(a)-(d) and the null hypothesis $H_0$, we have
\begin{align}\label{Test10}
\underset{\upsilon\geq0}{\sup}\bigg| P\Big(\widetilde{\mathcal{G}}^*_T(K)\leq\upsilon\Big|X_1^T,Y_1^T\Big)-P\Big(\widetilde{\mathcal{G}}_T(K)\leq\upsilon\Big)\bigg|=o_p(1),
\end{align}
where $X_1^T=\{X_1,\ldots,X_T\}$ and $Y_1^T=\{Y_1,\ldots,Y_T\}$.
\end{theorem}

\begin{remark}{\rm  If the $U_t$'s are assumed to be i.i.d, then the condition $\mathcal{C}_{N,T,k}= o(T^{1/3})$ in Theorems~\ref{theo1} and \ref{theo2} can be relaxed to ${k(\log NT/k)}^{7/3}=o(T^{1/3})$. However, if we allow the $U_t$'s to be alpha-mixing in Theorems~\ref{theo1} and \ref{theo2}, a more restricted condition on the order of $k$ relative to $T$ is required.}
\end{remark}

It is important to note that our theoretical results allow for the number of factors $r$ to grow slowly with $N$ and $T$. However, the restrictions on $k$ and $K$ in Theorems~\ref{theo1}-\ref{theo2} cannot be easily relaxed due to the use of the Gaussian approximation theory in \cite{Chang2021}. Nevertheless, our numerical results demonstrate that the simulation-based approach performs reasonably well even for large values of $K$.

\subsection{Power analysis}
Below we study the power properties of the proposed testing
procedures. To proceed, we impose the following conditions.
\begin{ass}\label{assum-2}{\rm~
\begin{itemize}
\item [\rm (a)]
$\max_{j_1\leq\dots\leq j_k} \sum_{i=1}^k \gamma_{j_i,j_i} \alpha^2_{j_i} \geq (2k + \delta)\log(N)$ for some $\delta>0$.
\item [\rm (b)] $|\boldsymbol{\alpha}|_0=\lceil N^\kappa\rceil$ for some $0 \leq \kappa <1/4$ and the non-zero locations are
randomly and uniformly drawn from $\{1, 2,\ldots, N\}$.% And the scheme is independent of $\{U_t-\Theta\}_{t=1}^T$.
\item [\rm (c)] Let $\text{diag}^{-1/2}(\Gamma)\Gamma\text{diag}^{-1/2}(\Gamma)=(\phi_{i,j})^{N}_{i,j=1}$. Assume that $\max_{1\leq i<j\leq N}|\phi_{i,j}|\leq c_0<1$.
\end{itemize}
}
\end{ass}

Define
$\widetilde{C}^*_\alpha=\inf\{\upsilon>0: P(\widetilde{G}^*_T(k)\leq \upsilon|X_1^T,Y_1^T)\geq1-\alpha\}$ as the simulation-based critical value.
Let $\widetilde{\mathcal{C}}_\alpha^*$
be the analogous quantity defined based on $\widetilde{\mathcal{G}}^*_T(K)$. The following theorem establishes the consistency of the testing procedures.

\begin{theorem}\label{theo3} Assume that $\mathcal{C}_{N,T,k}= o(T^{1/3})$ and $k^2(r^3+r^2m)\big(\log NT\big)^{5/2}/\sqrt{T}=o(1)$. Then under Assumptions \ref{assum-1} and \ref{assum-2} (a) (b) (c), we have
\begin{align}\label{Test110}
P(\widetilde{G}_T(k)>\widetilde{C}^*_\alpha)\rightarrow 1.
\end{align}
Moreover, suppose Assumption~(a) holds by replacing $k$ with $K$ and
$K^4(r^2+rm)\big(\log NT\big)^{5/2}/\sqrt{T} = o(1)$. Then we have
\begin{align}\label{Test120}
P(\widetilde{\mathcal{G}}_T(K)>\widetilde{\mathcal{C}}^*_\alpha)\rightarrow 1.
\end{align}
\end{theorem}

\section{Numerical studies}
\subsection{Competing methods}
We carry out simulation experiments to evaluate the finite sample performance of our tests. As a comparison, we also consider the FLY test of \cite{Fan2015}, the PY test of \cite{Pesaran2017}, and the testing procedures recently proposed by \cite{Feng2021}.
Specifically, we consider the adjusted Wald-statistic defined as
$$\mathcal{T}_{\text{FLY}}=\frac{T(1-\bar{X}^\top\mathbf{\widetilde{W}})
\widehat{\boldsymbol{\alpha}}^\top{\widehat{\Sigma^\mathrm{T}}}^{-1}\widehat{\boldsymbol{\alpha}}-\frac{N(T-r-1)}{T-r-3}}
{\frac{T-r-1}{T-r-3}\sqrt{\frac{2N(T-r-2)}{T-r-5}\big(1+(N-1)\tilde{\rho}^2_{N,T}\big)}},$$
where
$\mathbf{\widetilde{W}}=\big(\frac{1}{T}\sum_{t=1}^TX_tX^\top_t\big)^{-1}\bar{X}\in\mathbb{R}^{r\times 1}$,
and $\widehat{\Sigma^\mathrm{T}}$ is a thresholding covariance estimator [see e.g., \cite{Antoniadis2001}, \cite{Rothman2009}].

\cite{Pesaran2012,Pesaran2017} proposed the modified Wald-type test:
\begin{align*}
\mathcal{T}_{\text{PY}}
=\frac{N^{-1/2}\sum_{i=1}^N\Big(t^2_i-\frac{T-r-1}{T-r-3}\Big)}{\frac{T-r-1}{T-r-3}
\sqrt{\frac{2(T-r-2)}{T-r-5}\big(1+(N-1)\tilde{\rho}^2_{N,T}\big)}},
\end{align*}
where
\begin{align*}
&t^2_i=\frac{\widehat{\alpha}^2_i \big(\mathbf{1}^\top \mathbf{M}_\mathbf{X}\mathbf{1}\big)}{\widehat{U}^\top_{i\cdot}\widehat{U}_{i\cdot}/(T-r-1)}, \quad\widehat{V}_U=\sum_{t=1}^T\widehat{U}_t\widehat{U}_t^\top/(T-r-1),\\
&\tilde{\rho}^2_{N,T}=\frac{2}{N(N-1)}\sum_{i=2}^N\sum_{j=1}^{i-1}\hat{\rho}^2_{ij}
I\Big((T-r-1)\hat{\rho}^2_{ij}\geq \varrho_{N,\alpha}\Big),
\end{align*}
with $\mathbf{Y}_{i\cdot} = (y_{i1},\ldots,y_{iT})^\top$,
$\widehat{U}_{i\cdot}=(\widehat{u}_{i1},\ldots,\widehat{u}_{iT})^\top =\mathbf{M}_\mathbf{X}\big(\mathbf{Y}_{i\cdot}- \widehat{\alpha}_i\mathbf{1}\big)$, $\hat{\rho}_{ij}=\frac{\widehat{U}^\top_{i\cdot}\widehat{U}_{j\cdot}}{\sqrt{\widehat{U}^\top_{i\cdot}\widehat{U}_{i\cdot}}\sqrt{\widehat{U}^\top_{j\cdot}\widehat{U}_{j\cdot}}}$,
$\widehat{U}_t = Y_t - \widehat{\boldsymbol{\alpha}}^\top-\widehat{\boldsymbol{\beta}}X_t$ and
$\sqrt{\varrho_{N,\alpha}}=\Phi^{-1}(1-p_N/2)$. Here $\Phi$ is the standard normal distribution function and $p_N=0.1/(N-1)$.
$\widehat{U}_\cdot=(\widehat{u}_{i1},\ldots,\widehat{u}_{iT})^\top =\mathbf{M}_\mathbf{X}\big(\mathbf{Y}_{i\cdot}- \widehat{\alpha}_i\mathbf{1}\big)$
\cite{Feng2021} found that the maximum type statistic [\cite{Gungor2016}] defined as
\begin{align*}
\mathcal{T}_{\text{MAX}}=\max_{1\leq i\leq N} t^2_i.
\end{align*}
follows an extreme value distribution. Moreover, \cite{Feng2021} combined the $\mathcal{T}_{\text{PY}}$ and $\mathcal{T}_{\text{MAX}}$ tests to obtain a combination test $\mathcal{T}_{\text{COM}}$ that is capable of detecting both sparse and dense alternatives. See \cite{Feng2021} for the details.

\subsection{Simulation results}
Following the simulation studies of \cite{Pesaran2017} and \cite{Feng2021},
the experiments are designed to mimic the commonly used Fama-French three-factor model, where the factors
$X_t$ have strong serial correlation and heterogeneous variance. In particular, we consider the model
\begin{align*}
y_{it} = \alpha_i + \beta^\top_i X_t + u_{it},\quad i=1,2,\dots,N,\quad t=1,2,\dots,T.
\end{align*}
Here $\beta_i=(\beta_{i1}, \beta_{i2}, \beta_{i3})^\top$ for $1\leq i\leq N$ and $X_t=(x_{1t}, x_{2t}, x_{3t})^\top$ for $1\leq t\leq T$, which corresponds to the three factors in the Fama-French model, namely the market factor, the small minus big (SMB) and high minus low (HML). Each factor follows a GARCH(1, 1) process,
and all the coefficients are the same as those in \cite{Pesaran2017}. Specifically,
\begin{align*}
&\textmd{Market factor:}~~x_{1t} =0.53 + 0.06x_{1,t-1} + h_{1t}^{1/2}\zeta_{1t},\\
&\textmd{SMB factor:}~~x_{2t} =0.19 + 0.19x_{2,t-1} + h_{2t}^{1/2}\zeta_{2t}, \\
&\textmd{HML factor:}~~x_{3t} =0.19 + 0.05x_{3,t-1} + h_{3t}^{1/2}\zeta_{3t},
\end{align*}
where $\zeta_{jt}'s$ are simulated from a standard normal distribution,
and the variance terms $h_{jt}$ are generated from
\begin{align*}
&\textmd{Market factor:}~~h_{1t} =0.89 + 0.85h_{1,t-1} + 0.11h_{1t-1}\zeta^2_{1t-1},\\
&\textmd{SMB factor:}~~h_{2t} =0.62 + 0.74h_{2,t-1} + 0.19h_{2t-1}\zeta^2_{2t-1},\\
&\textmd{HML factor:}~~h_{3t} =0.80 + 0.76h_{3,t-1} + 0.15h_{3t-1}\zeta^2_{3t-1}.
\end{align*}
Similar to \cite{Pesaran2017} and \cite{Feng2021}, the above process is generated over the periods $t =-49, \ldots, 0, 1, \ldots, T$ with
 $x_{j,-50}= 0$ and $h_{j,-50} = 1$ for any $j=1, 2$ and $3$. We discard the first 50 observations and use the data from time 1 to time $T$ to evaluate the performance of the tests.

Following \cite{Pesaran2017} and \cite{Feng2021}, to capture the main features of the individual asset returns and their cross-correlations, we set the idiosyncratic errors to be $$U_t=\Sigma^{1/2}_U\varepsilon_t,$$
and consider three types of distribution for $\varepsilon_t$ as follows
\begin{itemize}
\item [\rm (1)]
$\varepsilon_t \sim N(\textbf{0},\mathrm{I}_N)$, $t=1,2, \ldots, T$;
\item [\rm (2)] $\varepsilon_{it} \overset{iid}{\sim} t(3)/\sqrt{3}$, $i=1, 2, \ldots, N$, $t=1, 2, \ldots, T$;
\item [\rm (3)] $\varepsilon_{t}$ consists of $N$ independent ARCH processes, i.e., each
component process is of the form $\varepsilon_{i,t} = \sigma_{i,t}e_{i,t}$, where the $e_{i,t}$ are independent random variables following $N(0, 1)$ or $t(3)/\sqrt{3}$, and
$\sigma_{i,t}=\gamma_0+\gamma_1\varepsilon^2_{i,t-1}$ with $\gamma_0$ and $\gamma_1$ generated from, respectively, the uniform distributions on $(0.25, 0.5)$ and $(0, 0.5)$
independently for different component processes.
\end{itemize}
Meanwhile, the factor loadings $\beta_{i1}$, $\beta_{i2}$ and $\beta_{i3}$ are generated independently
from the uniform distributions on $(0.2, 2), (-1, 1.5)$ and $(-1.5, 1.5)$, respectively.

\begin{figure}
	\centering
	\includegraphics[height=6cm,width=15cm]{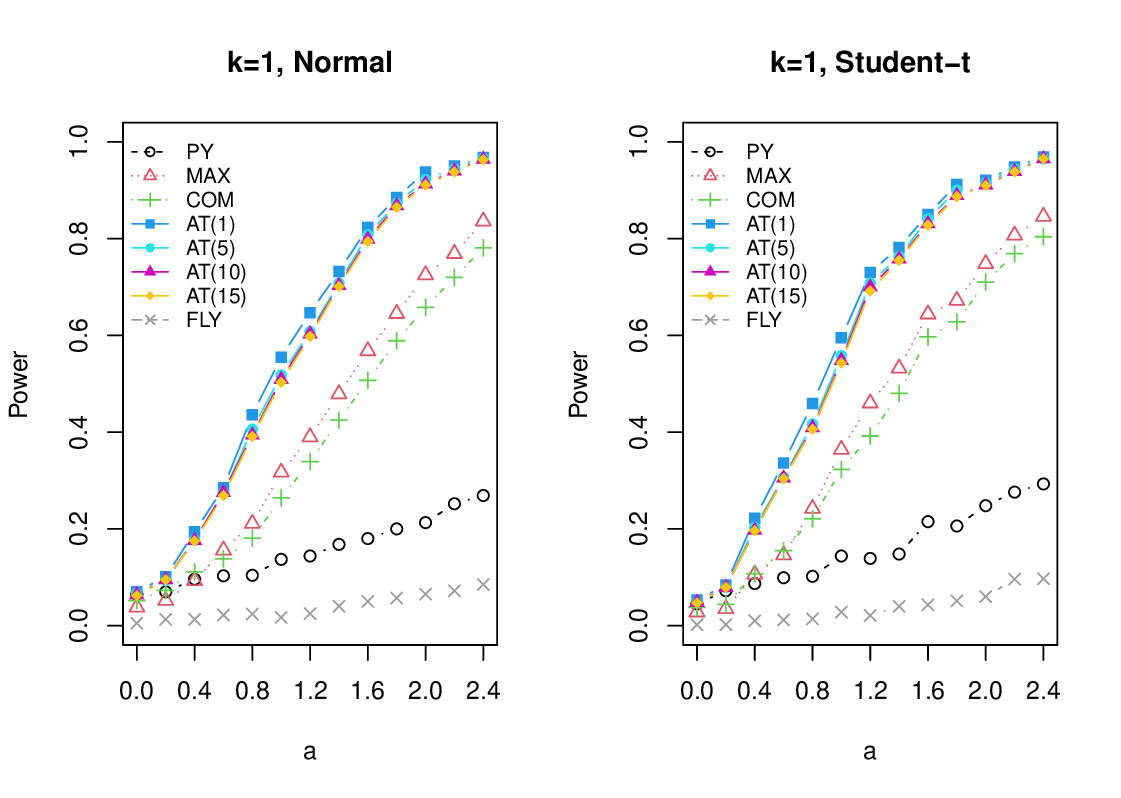}
    \includegraphics[height=6cm,width=15cm]{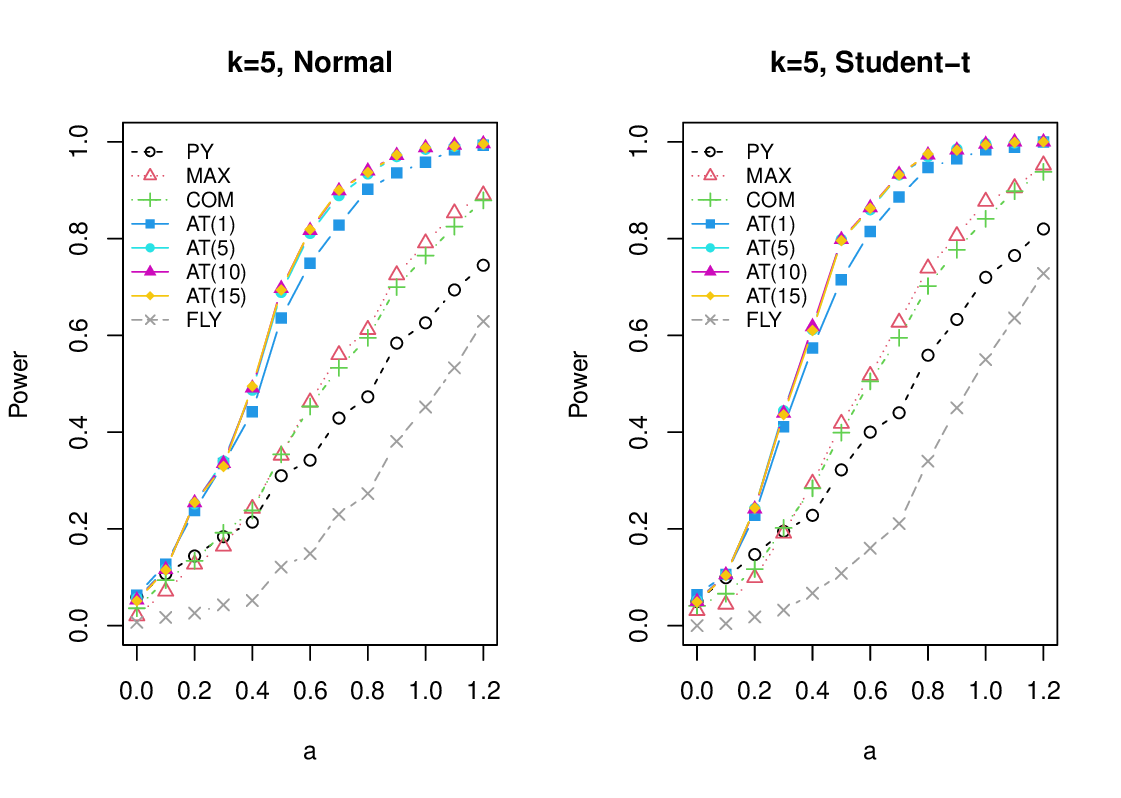}
    \includegraphics[height=6cm,width=15cm]{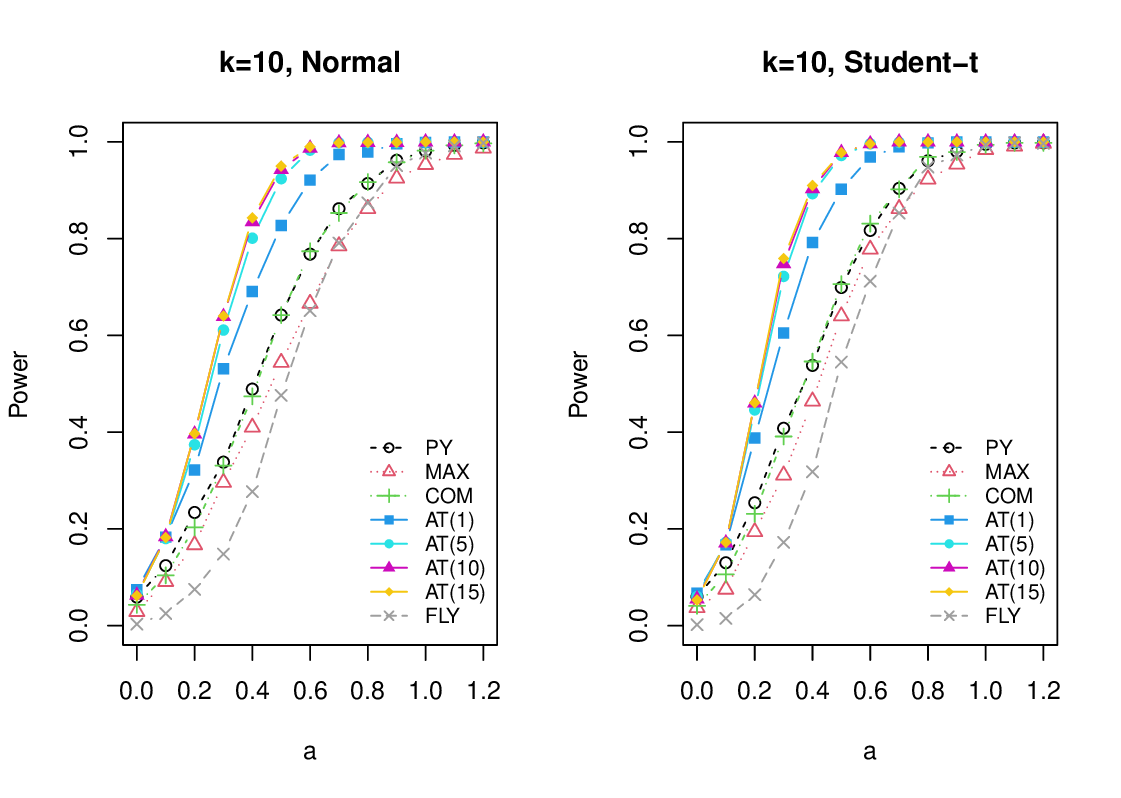}
    \includegraphics[height=6cm,width=15cm]{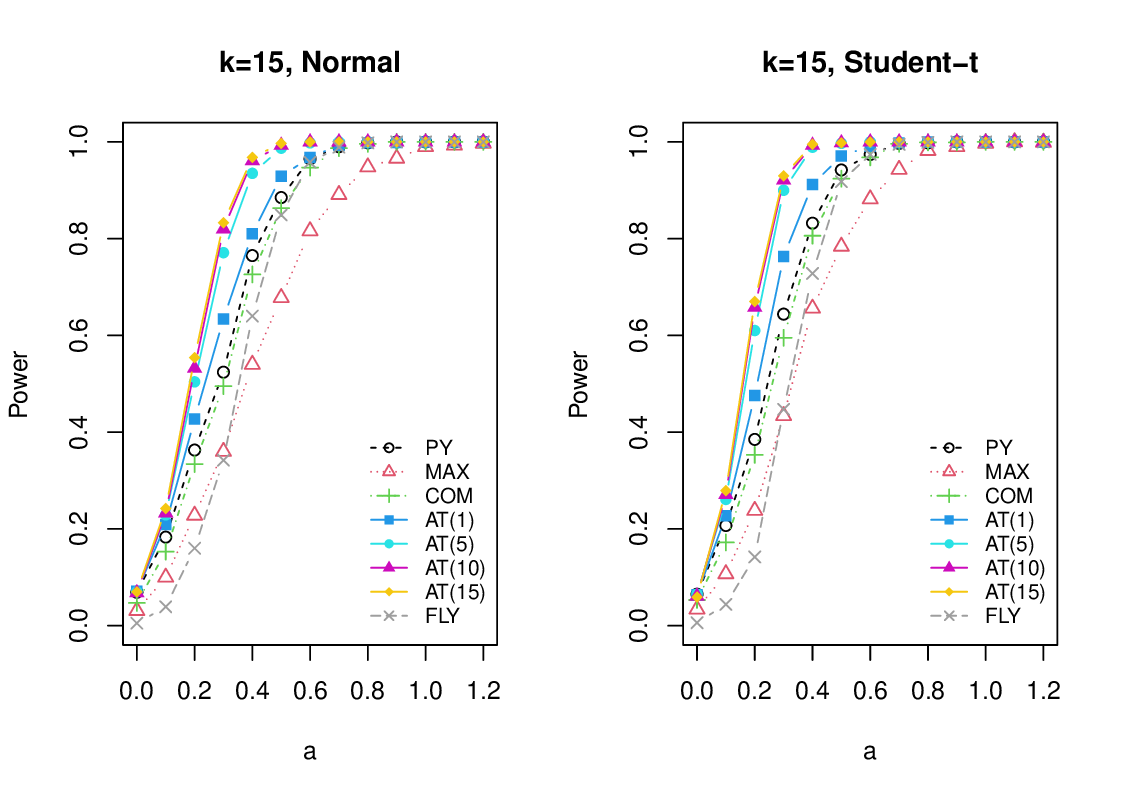}
	\caption{Power curves for different tests, where $k = 1, 5, 10, 15$ with $T=100, N = 100$, and $\Sigma_U=\big(\sigma_{ij}\big)_{i,j=1}^N$ with $\sigma_{ij}=0.6^{|i-j|}$. ``AT" represents the proposed test statistics. }  	\label{FIG:0}
\end{figure}

To better understand the power performance of all the methods under different sparsity levels, we
first present a small numerical study to investigate the power behavior when $k=1, 5, 10$, and $15$, and $T=100, N=100$.
We focus on the AR(1) covariance structure $\Sigma_U=\big(\sigma_{ij}\big)_{i,j=1}^N$ with $\sigma_{ij}=0.6^{|i-j|}$ and $\Gamma=\Sigma^{-1}_U$.
The non-zero entries of $\boldsymbol{\alpha}$
are equal to $\alpha_i=\omega_i\sqrt{a\log(N)/T}$, where $\omega_i$ are i.i.d. random variables with
$P(\omega_i=\pm1)=1/2$. %and $a=0.0, 0.2, 0.4, 0.6, 0.8, 1.0, 1.2, 1.4$.
Figure~\ref{FIG:0} shows the power curves as a function of the signal strength parameter $a$. The results clearly indicate that the proposed test outperforms the other tests in all cases considered. In particular, the COM, PY, and FLY tests are more sensitive to dense signals, while the MAX test performs better than these tests for sparse signals. Notably, the power of $\widetilde{\mathcal{G}}_T(K)$ with $K>1$ is higher than that of $\widetilde{\mathcal{G}}_T(1)$, suggesting that it is beneficial to choose a larger value of $K$ in practice. It should be noted that while the restrictions on $k$ and $K$ in our theoretical results cannot be relaxed, the simulation-based approach performs reasonably well even when $K$ takes large values, as suggested by our numerical results.

Next, we consider the two cases for the error covariance $\Sigma_U$:
\begin{itemize}
\item [\rm (1)] In the first case, we assume $\Sigma_U=\Lambda^{1/2}R\Lambda^{1/2}$.
Similar to \cite{Pesaran2017} and \cite{Feng2021}, we let $\Lambda=\textmd{diag}(\sigma_{11},\cdots,\sigma_{NN})$ with the diagonal entries $\sigma_{ii}$ drawn rom the uniform distribution on (1, 2). We set the correlation matrix of $U_t$ to be
$$R = \mathrm{I}_N + \textbf{b}\textbf{b}^\top - \textmd{diag}(\textbf{b})^2,$$
where $\textbf{b} = (b_1,\ldots, b_N)^\top$.
The first and the last $\lfloor N^{\delta_\gamma}\rfloor$ with $\delta_\gamma (= 1/4, 1/2, 3/5)$ elements of $\textbf{b}$ are drawn from the uniform distribution on $(0.7, 0.9)$ to generate different degrees of error cross-sectional dependence, where $\delta_\gamma = 1/4$ corresponds to the case of weak correlation and $\delta_\gamma = 3/5$ represents strong correlation. The remaining elements are set to 0.

\item [\rm (2)] In the second case, we modify the example in Section 5.3 of \cite{Pesaran2017}. In particular,
$$\Sigma_U = \mathfrak{L}\mathfrak{L}^\top + (\mathrm{I}_N -\rho_\epsilon \mathcal{W})^{-1}(\mathrm{I}_N -\rho_\epsilon \mathcal{W}^\top)^{-1},$$
with $\mathfrak{L}=(\mathfrak{l}_1,\ldots,\mathfrak{l}_{\lfloor N^{\delta_\gamma}\rfloor},0,\ldots,0)^\top$.
Here $\mathfrak{l}_i$ is generated independently from the uniform distribution on (0.7, 0.9) for $i = 1, \ldots, \lfloor N^{\delta_\gamma}\rfloor$. Let $\rho_\epsilon=0.5$, and $\mathcal{W}=(w_{ij})_{N\times N}$. All elements in $\mathcal{W}$ are zero except that $w_{i+1,i}=w_{j-1,j} = 0.5$ for $i = 1, \ldots, N -2$ and $j = 3, \ldots, N$, and $w_{12}= w_{N,N-1} =1$.
\end{itemize}

We set $\boldsymbol{\alpha}=0$ under the null hypothesis. To evaluate the power of our tests and the other competing tests under the alternative hypothesis, we consider the two following scenarios.

\begin{itemize}
\item [\rm (1)] Scenario 1: $k = \lfloor N^{1/4}\rfloor$ and the non-zero entries of $\boldsymbol{\alpha}$
are equal to $\alpha_i=\omega_i\sqrt{2a\log(N)/T}$, where $\omega_i$ are i.i.d. random variables with
the Uniform distribution on (0, 1) and $a=1, 2, \ldots, 5$;
\item [\rm (2)] Scenario 2: $k = \lfloor N^{1/3}\rfloor$ and the non-zero entries of $\boldsymbol{\alpha}$
are equal to $\alpha_i=\omega_i\sqrt{2a\log(N)/T}$, where $\omega_i$ are i.i.d. random variables with
$P(\omega_i=\pm1)=1/2$ and $a=0.4, 0.8, 1.2, 1.6, 2.0$.
\end{itemize}

We present the empirical rejection rates for $\Sigma_U$ under the null in Case~(1) in Table~\ref{tab1}. Overall, all tests have reasonable empirical size, with the PY, MAX, and COM tests performing better than the other tests. When $\varepsilon_t$ follows the normal distribution and $N=100, 200$, the proposed adaptive test slightly over-rejects, while the other tests provide quite accurate control of the size. However, when $\delta_\gamma=1/2$ and $\delta_\gamma=3/5$, the FLY test is quite conservative.

\begin{figure}
	\centering
	\includegraphics[height=6cm,width=15cm]{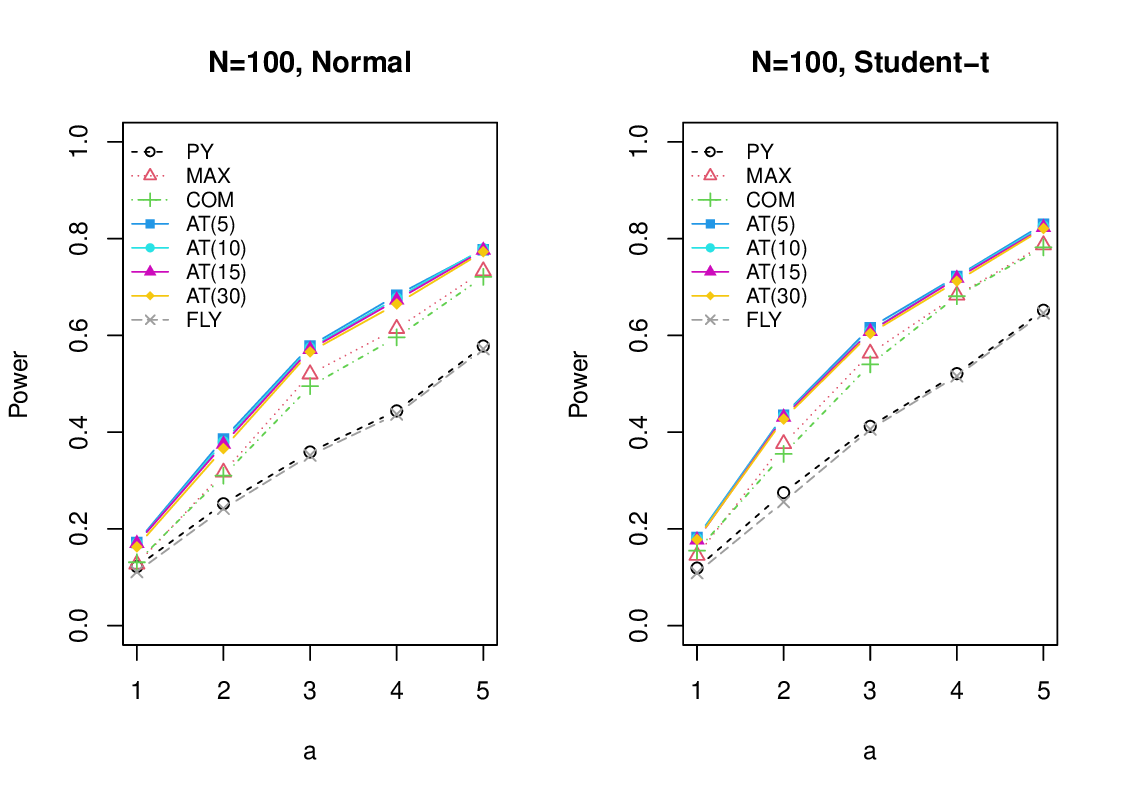}
	\caption{Power (\%) of different tests for $\Sigma_U$ in Case~(1) with $\delta_\gamma = 1/4$ and $k = \lfloor N^{1/4}\rfloor$}  	\label{FIG:1}
\end{figure}

\begin{figure}
	\centering
	\includegraphics[height=6cm,width=15cm]{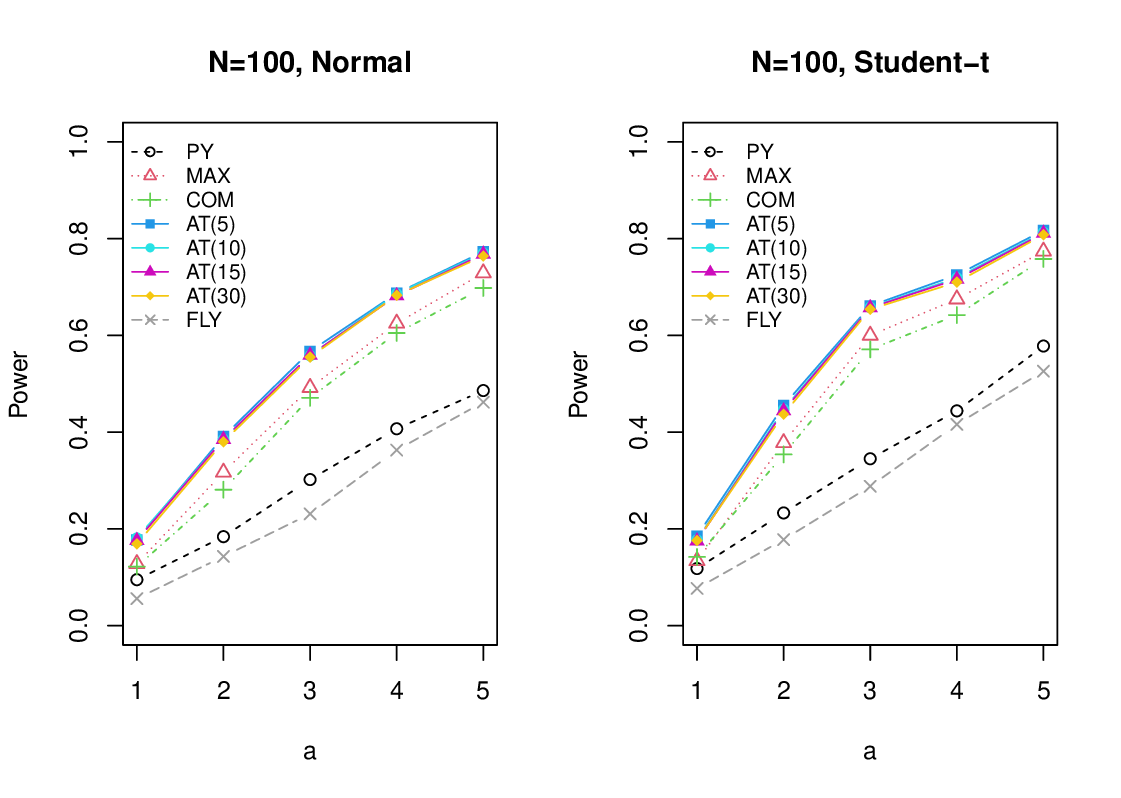}
	\caption{Power (\%) of different tests for $\Sigma_U$ in Case~(1) with $\delta_\gamma = 1/2$ and $k = \lfloor N^{1/4}\rfloor$}  	\label{FIG:2}
\end{figure}

\begin{figure}
	\centering
	\includegraphics[height=6cm,width=15cm]{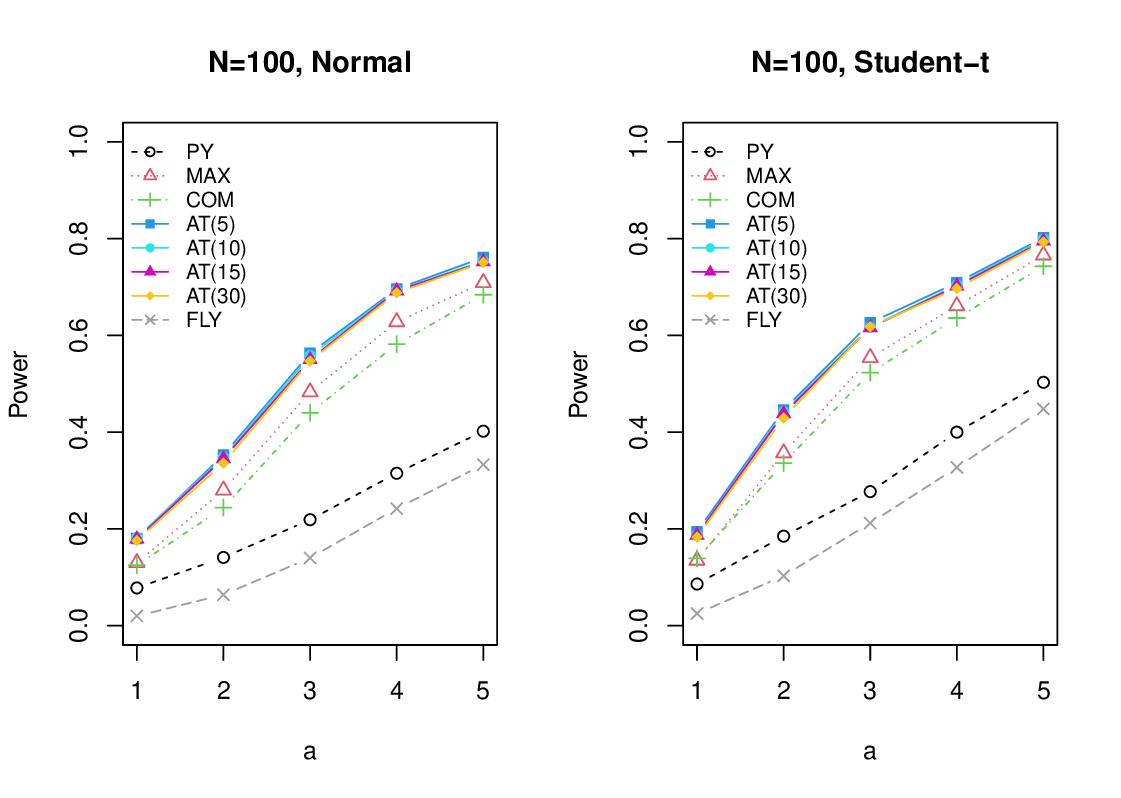}
	\caption{Power (\%) of different tests for $\Sigma_U$ in Case~(1) with $\delta_\gamma = 3/5$ and $k = \lfloor N^{1/4}\rfloor$}  	\label{FIG:3}
\end{figure}

\begin{figure}
	\centering
	\includegraphics[height=6cm,width=15cm]{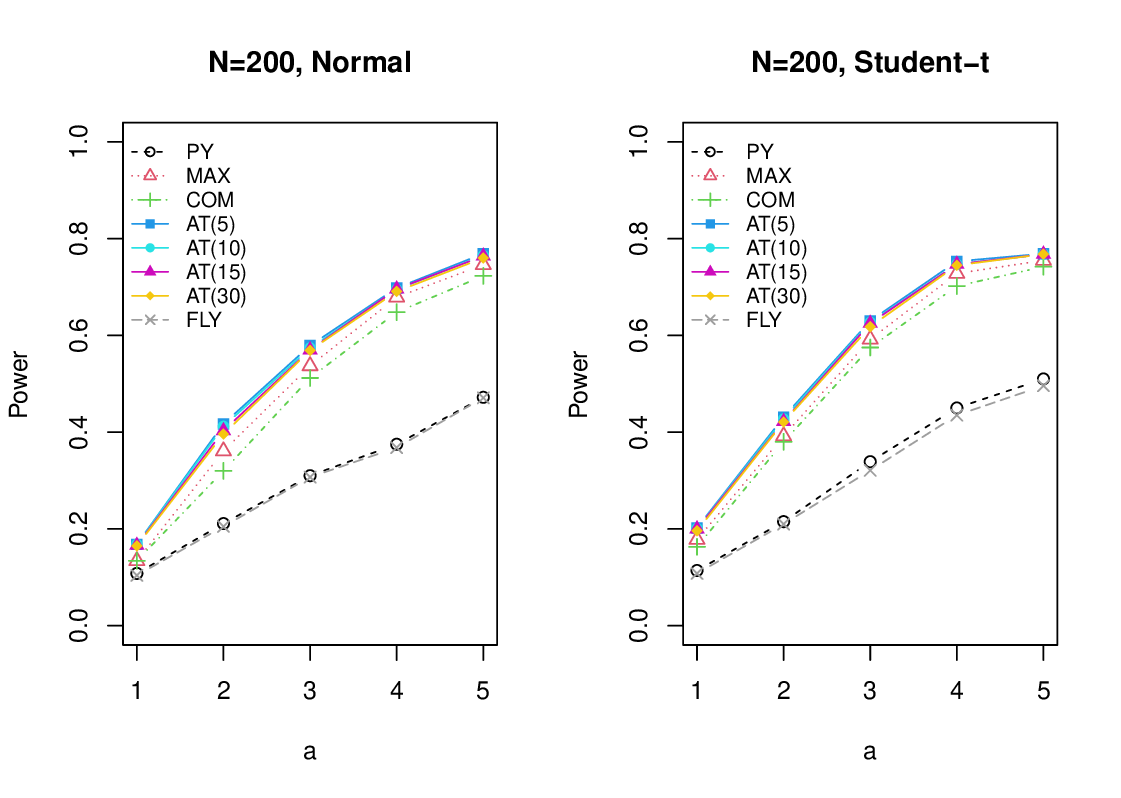}
	\caption{Power (\%) of different tests for $\Sigma_U$ in Case~(1) with $\delta_\gamma = 1/4$ and $k = \lfloor N^{1/4}\rfloor$}  	\label{FIG:4}
\end{figure}

\begin{figure}
	\centering
	\includegraphics[height=6cm,width=15cm]{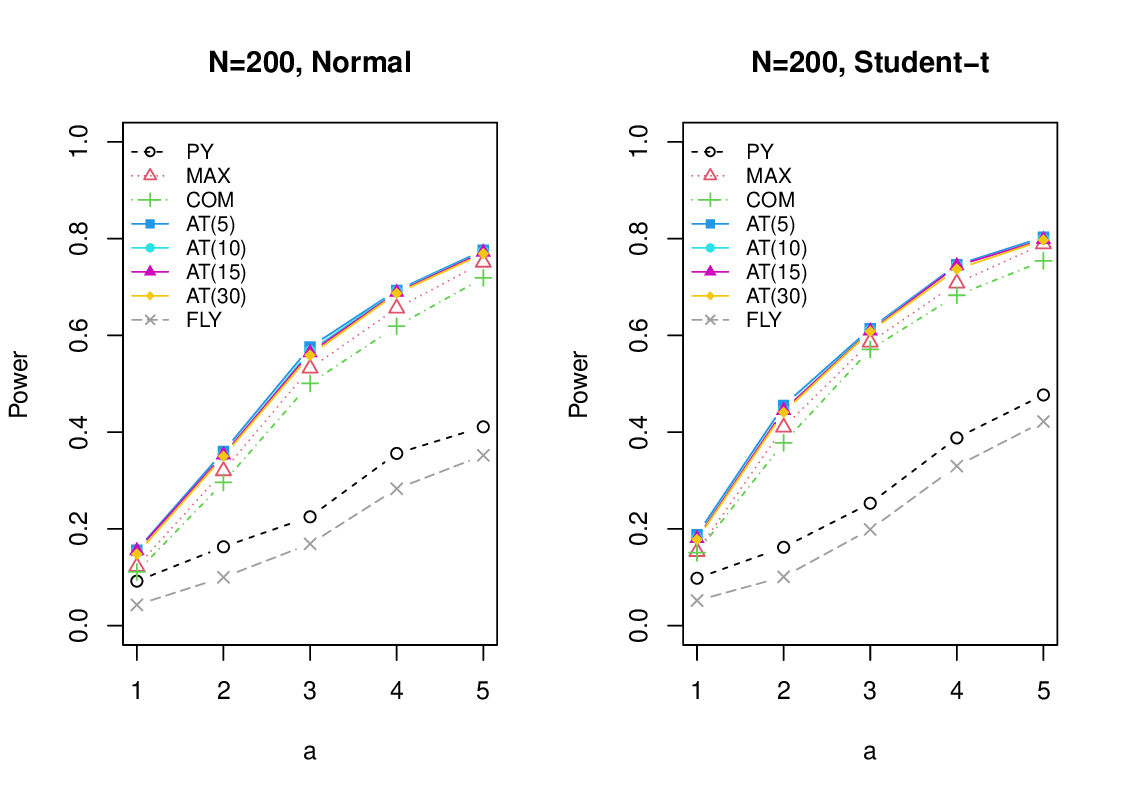}
	\caption{Power (\%) of different tests for $\Sigma_U$ in Case~(1) with $\delta_\gamma = 1/2$ and $k = \lfloor N^{1/4}\rfloor$}  	\label{FIG:5}
\end{figure}

\begin{figure}
	\centering
	\includegraphics[height=6cm,width=15cm]{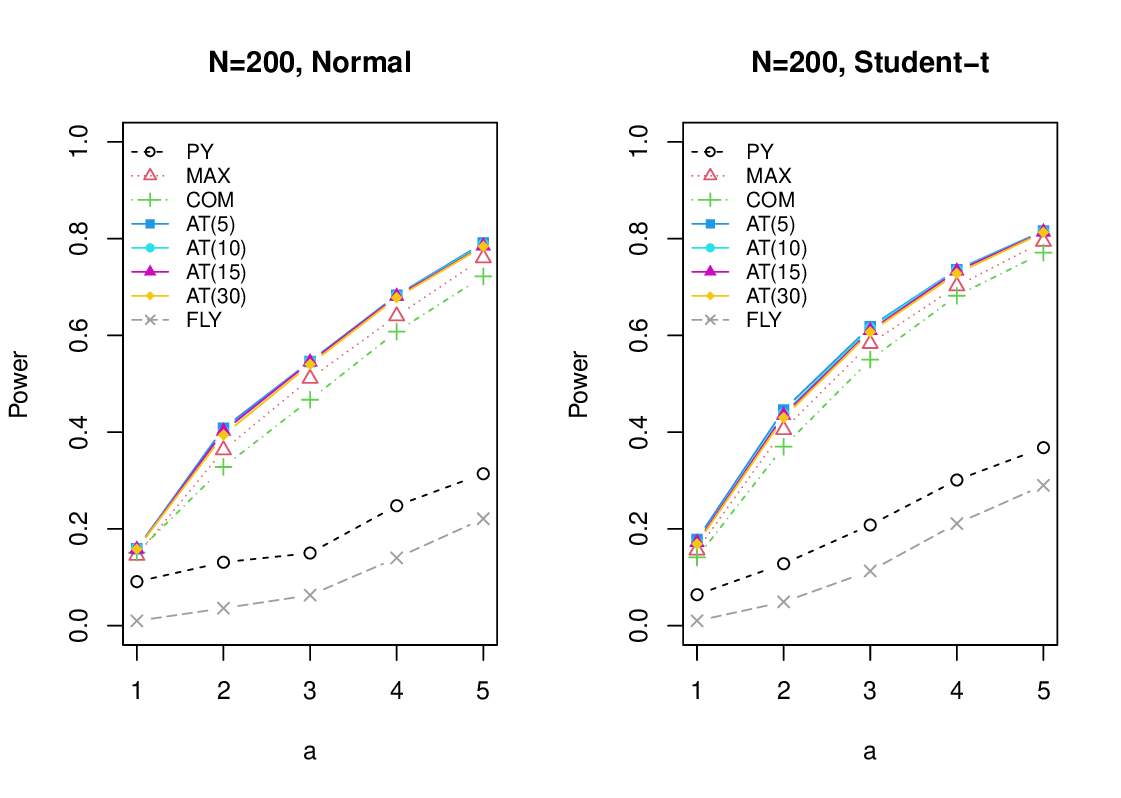}
	\caption{Power (\%) of different tests for $\Sigma_U$ in Case~(1) with $\delta_\gamma = 3/5$ and $k = \lfloor N^{1/4}\rfloor$}  	\label{FIG:6}
\end{figure}

\begin{figure}
	\centering
	\includegraphics[height=6cm,width=15cm]{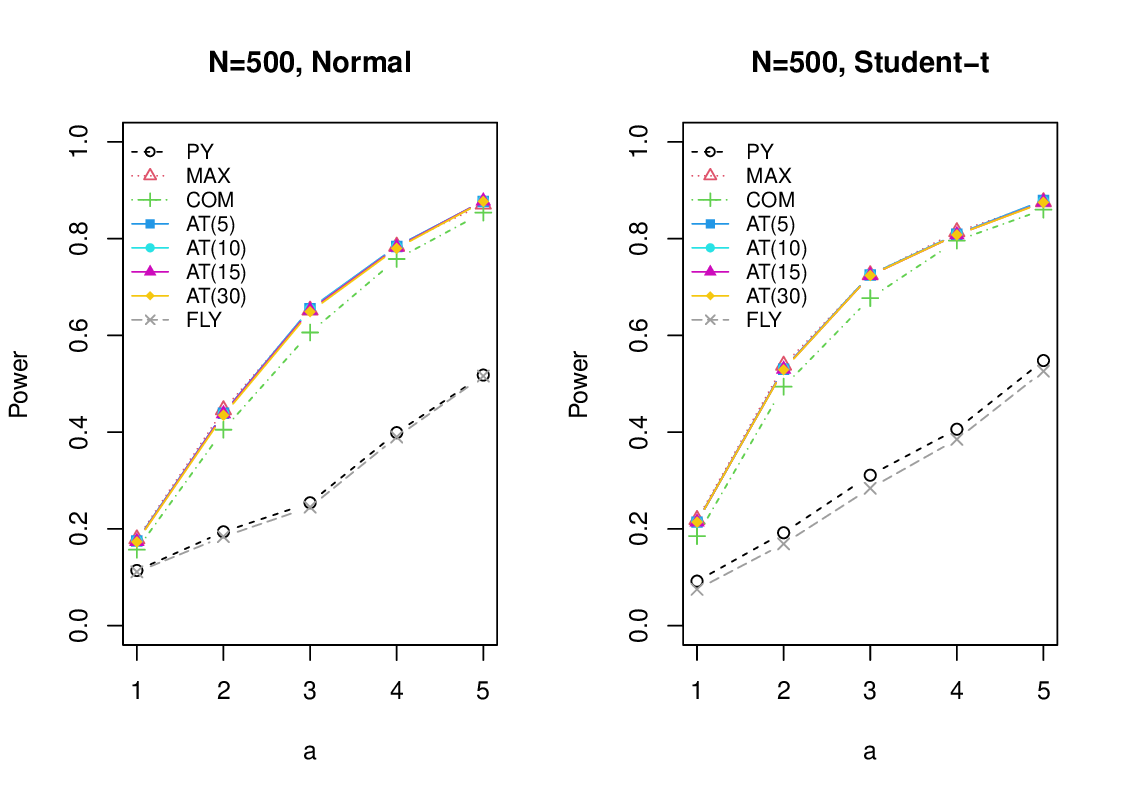}
	\caption{Power (\%) of different tests for $\Sigma_U$ in Case~(1) with $\delta_\gamma = 1/4$ and $k = \lfloor N^{1/4}\rfloor$}  	\label{FIG:7}
\end{figure}

\begin{figure}
	\centering
	\includegraphics[height=6cm,width=15cm]{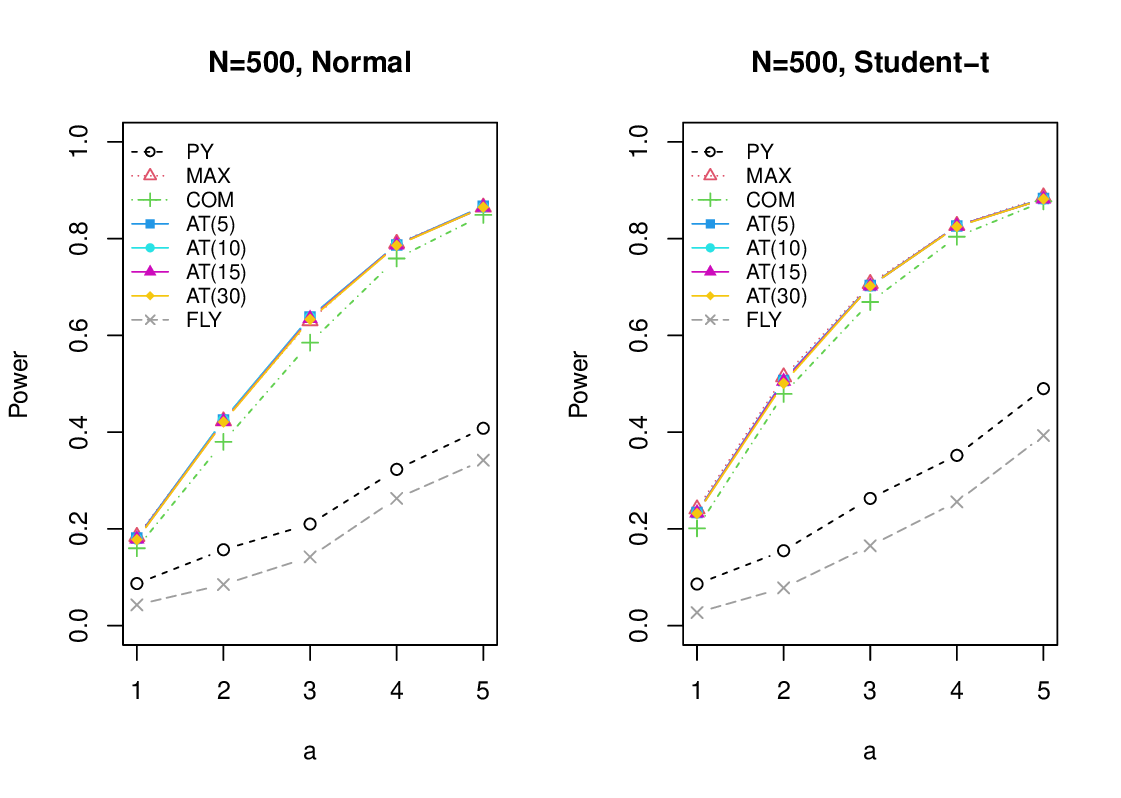}
	\caption{Power (\%) of different tests for $\Sigma_U$ in Case~(1) with $\delta_\gamma = 1/2$ and $k = \lfloor N^{1/4}\rfloor$}  	\label{FIG:8}
\end{figure}

\begin{figure}
	\centering
	\includegraphics[height=6cm,width=15cm]{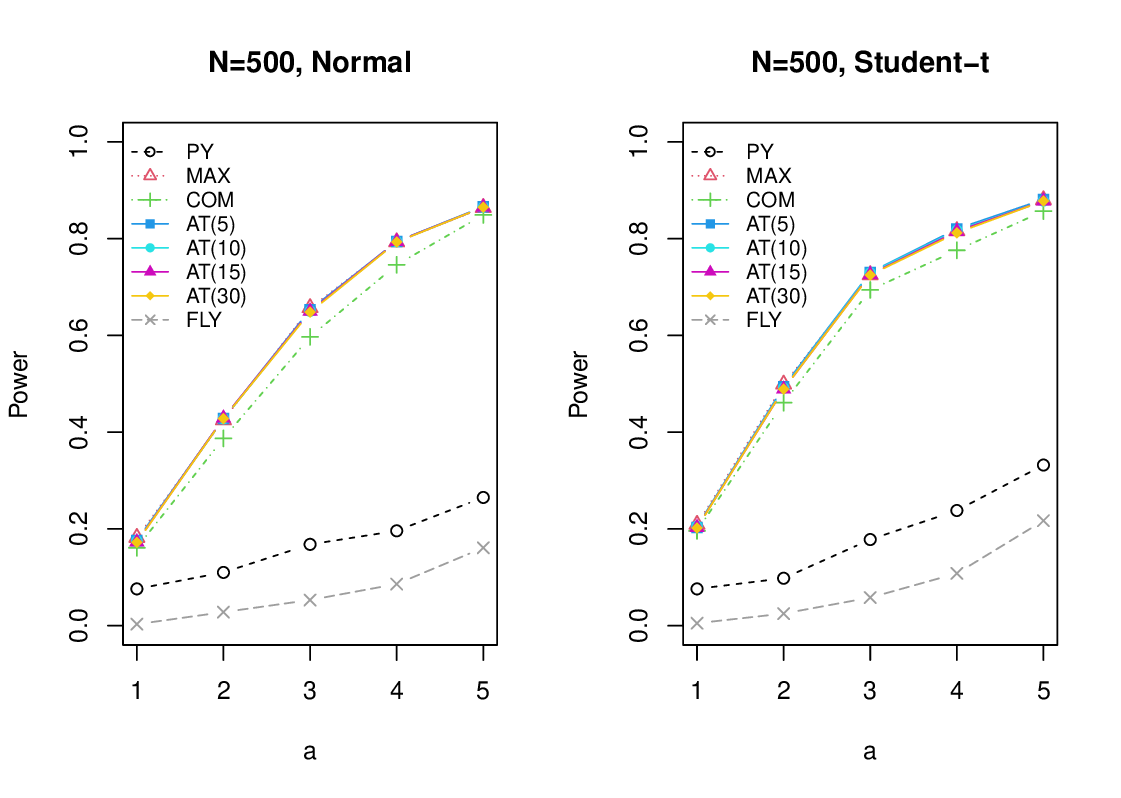}
	\caption{Power (\%) of different tests for $\Sigma_U$ in Case~(1) with $\delta_\gamma = 3/5$ and $k = \lfloor N^{1/4}\rfloor$}  	\label{FIG:9}
\end{figure}

Figures~\ref{FIG:1}-\ref{FIG:9} provide a summary of the rejection rates of different procedures under the alternatives for $\Sigma_U$ in Case~(1), with $k = \lfloor N^{1/4}\rfloor$. The proposed test, as well as the MAX and COM tests, outperform the PY and FLY tests in all cases, regardless of the values of $\delta_\gamma$. In the case of weak signals (i.e., $N=100, 200$), the proposed test delivers more power than the MAX and COM tests. As discussed in \cite{Feng2021}, the PY and FLY tests are sum-of-squares type tests that are sensitive to dense signals, and are likely to perform poorly in the case of sparse signals. Not surprisingly, strong cross-sectional dependence controlled by the parameter $\delta_\gamma$ harms the power of the PY and FLY tests. As $\delta_\gamma$ increases, the power of the PY and FLY tests decreases. Similar observations are found in the other cases; see Figures 1-9 for $\Sigma_U$ in Case(1) in the Supplementary material.

\begin{figure}
	\centering
	\includegraphics[height=6cm,width=15cm]{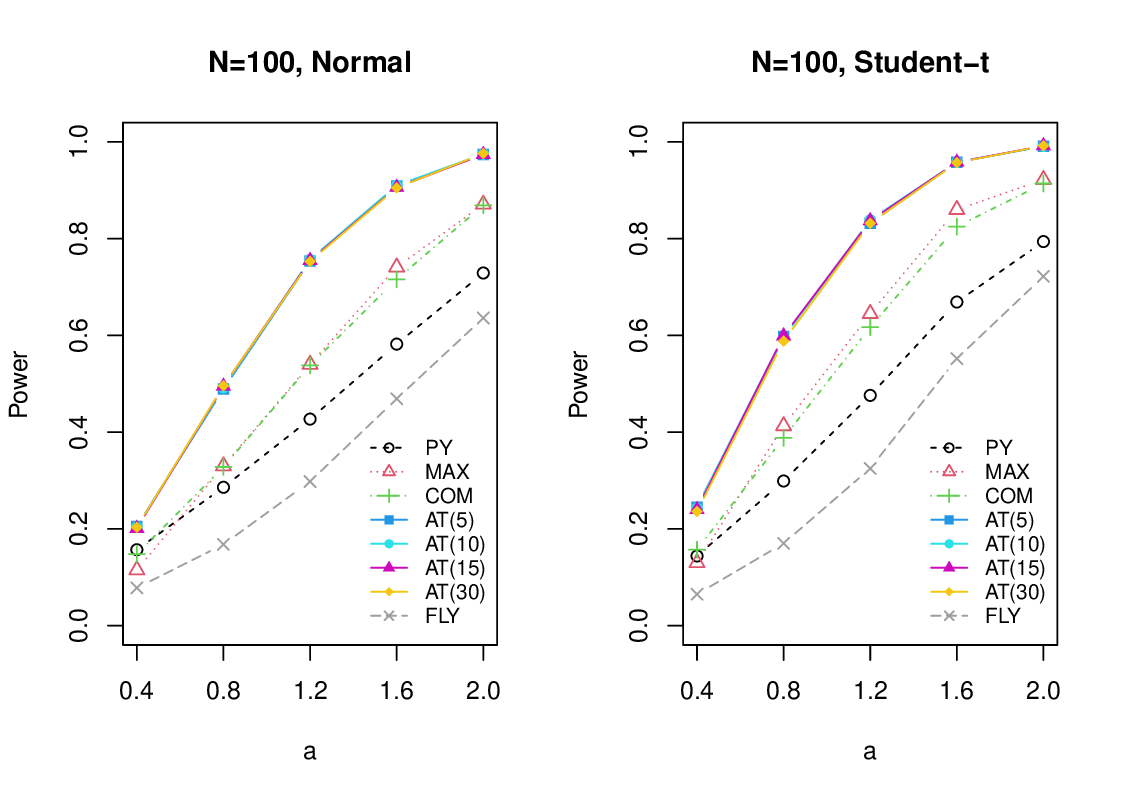}
	\caption{Power (\%) of different tests for $\Sigma_U$ in Case~(2) with $\delta_\gamma = 1/4$ and $k = \lfloor N^{1/3}\rfloor$}  	\label{FIG:10}
\end{figure}

\begin{figure}
	\centering
	\includegraphics[height=6cm,width=15cm]{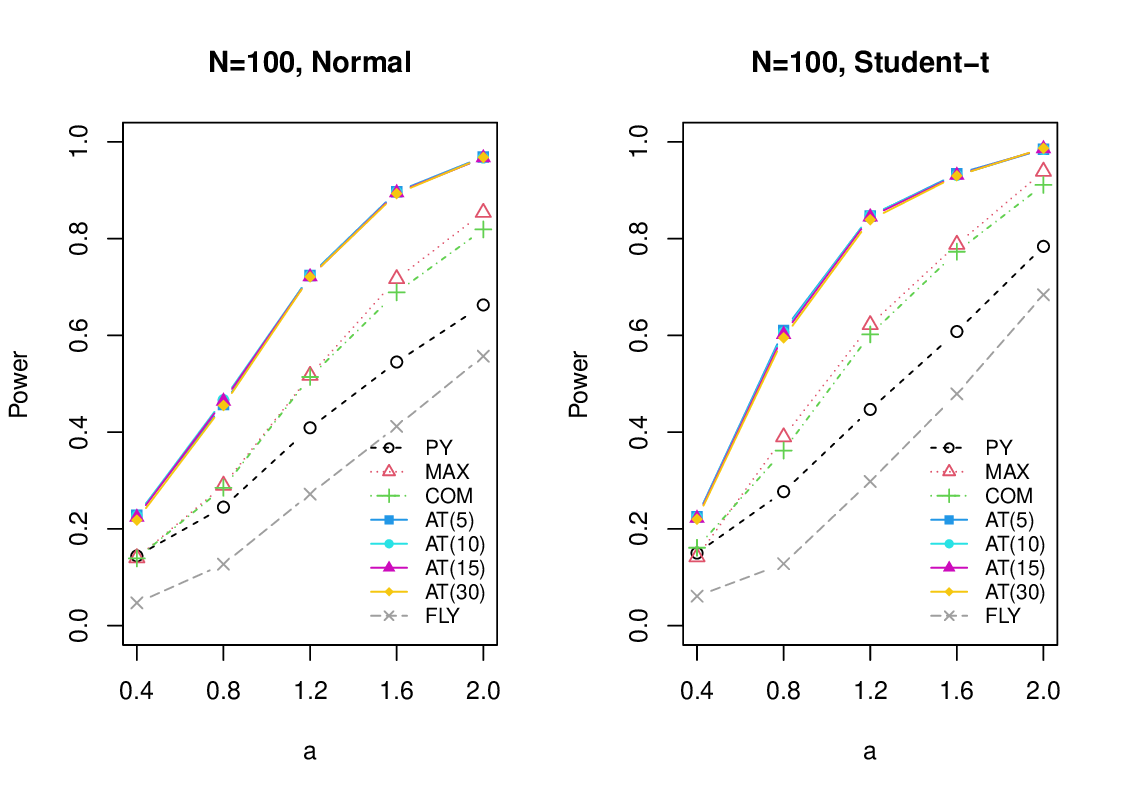}
	\caption{Power (\%) of different tests for $\Sigma_U$ in Case~(2) with $\delta_\gamma = 1/2$ and $k = \lfloor N^{1/3}\rfloor$}  	\label{FIG:11}
\end{figure}

\begin{figure}
	\centering
	\includegraphics[height=6cm,width=15cm]{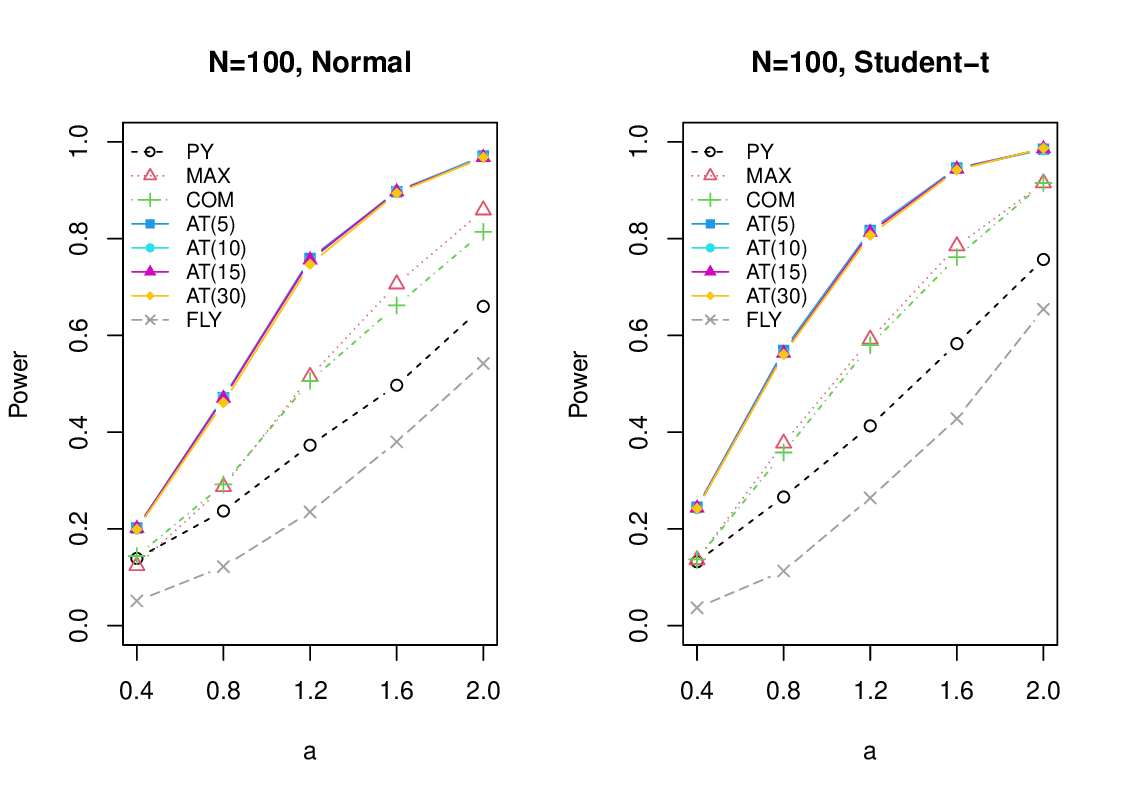}
	\caption{Power (\%) of different tests for $\Sigma_U$ in Case~(2) with $\delta_\gamma = 3/5$ and $k = \lfloor N^{1/3}\rfloor$}  	\label{FIG:12}
\end{figure}

\begin{figure}
	\centering
	\includegraphics[height=6cm,width=15cm]{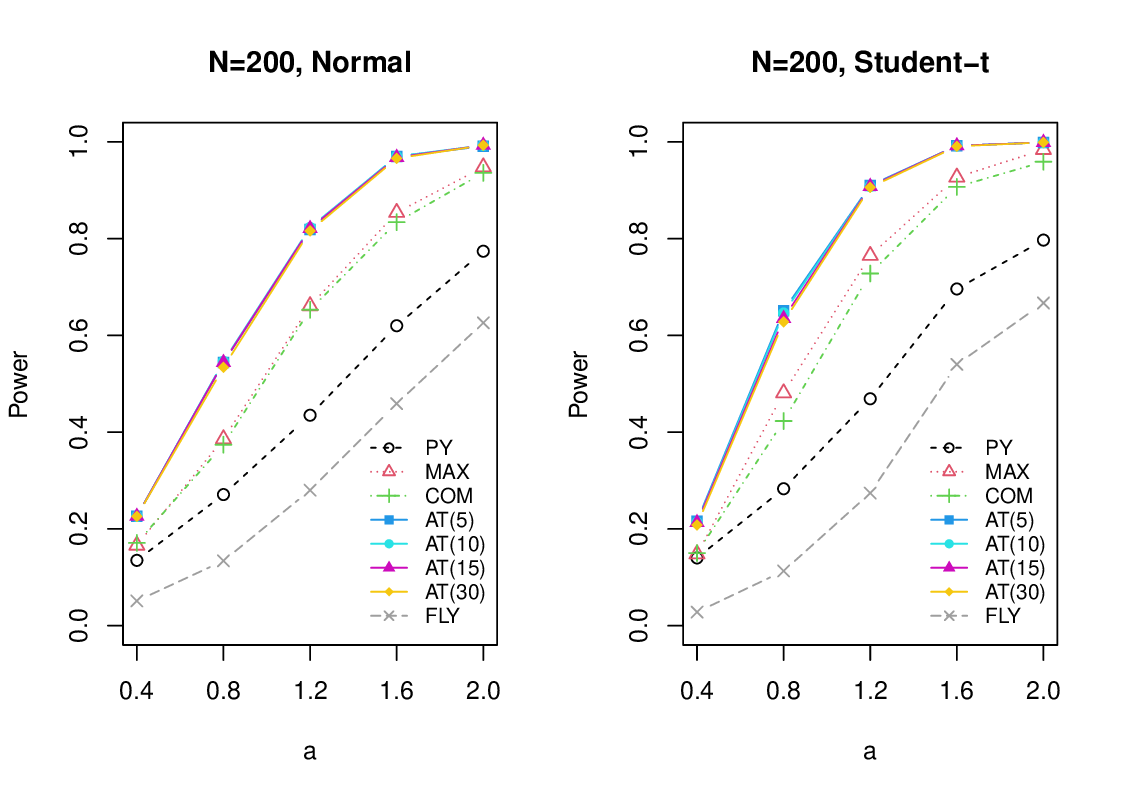}
	\caption{Power (\%) of different tests for $\Sigma_U$ in Case~(2) with $\delta_\gamma = 1/4$ and $k = \lfloor N^{1/3}\rfloor$}  	\label{FIG:13}
\end{figure}

\begin{figure}
	\centering
	\includegraphics[height=6cm,width=15cm]{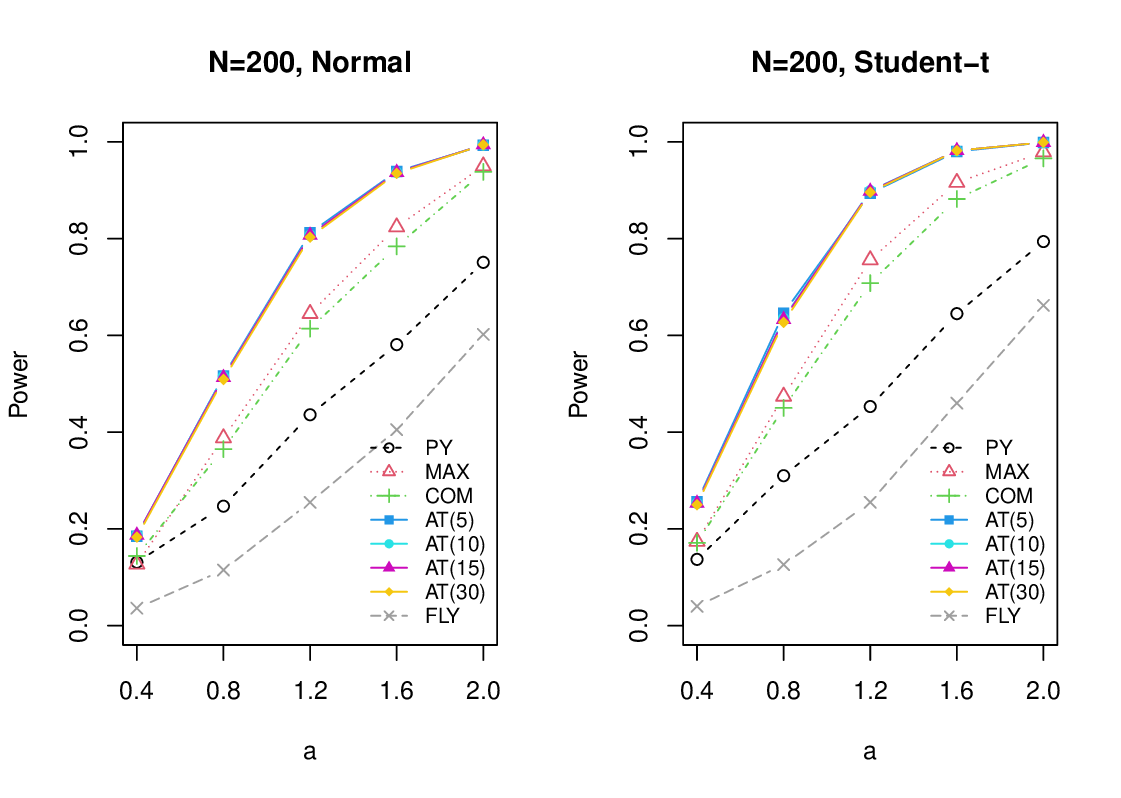}
	\caption{Power (\%) of different tests for $\Sigma_U$ in Case~(2) with $\delta_\gamma = 1/2$ and $k = \lfloor N^{1/3}\rfloor$}  	\label{FIG:14}
\end{figure}

\begin{figure}
	\centering
	\includegraphics[height=6cm,width=15cm]{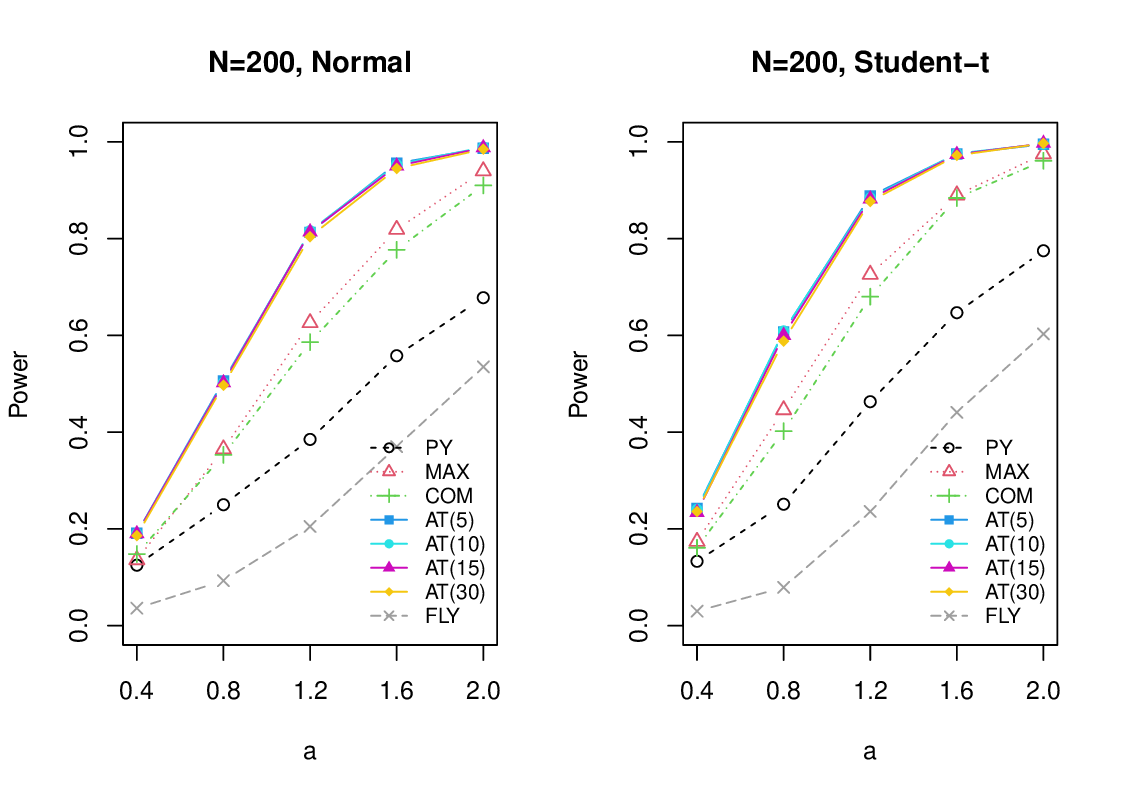}
	\caption{Power (\%) of different tests for $\Sigma_U$ in Case~(2) with $\delta_\gamma = 3/5$ and $k = \lfloor N^{1/3}\rfloor$}  	\label{FIG:15}
\end{figure}

\begin{figure}
	\centering
	\includegraphics[height=6cm,width=15cm]{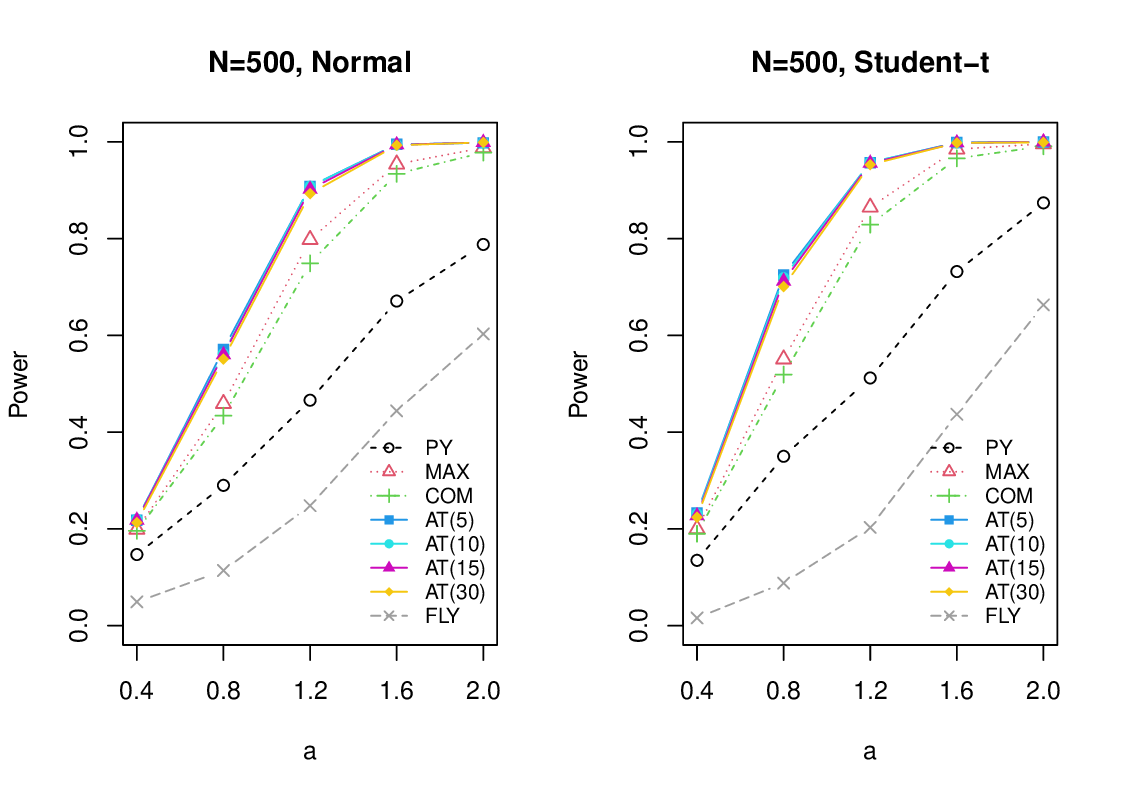}
	\caption{Power (\%) of different tests for $\Sigma_U$ in Case~(2) with $\delta_\gamma = 1/4$ and $k = \lfloor N^{1/3}\rfloor$}  	\label{FIG:16}
\end{figure}

\begin{figure}
	\centering
	\includegraphics[height=6cm,width=15cm]{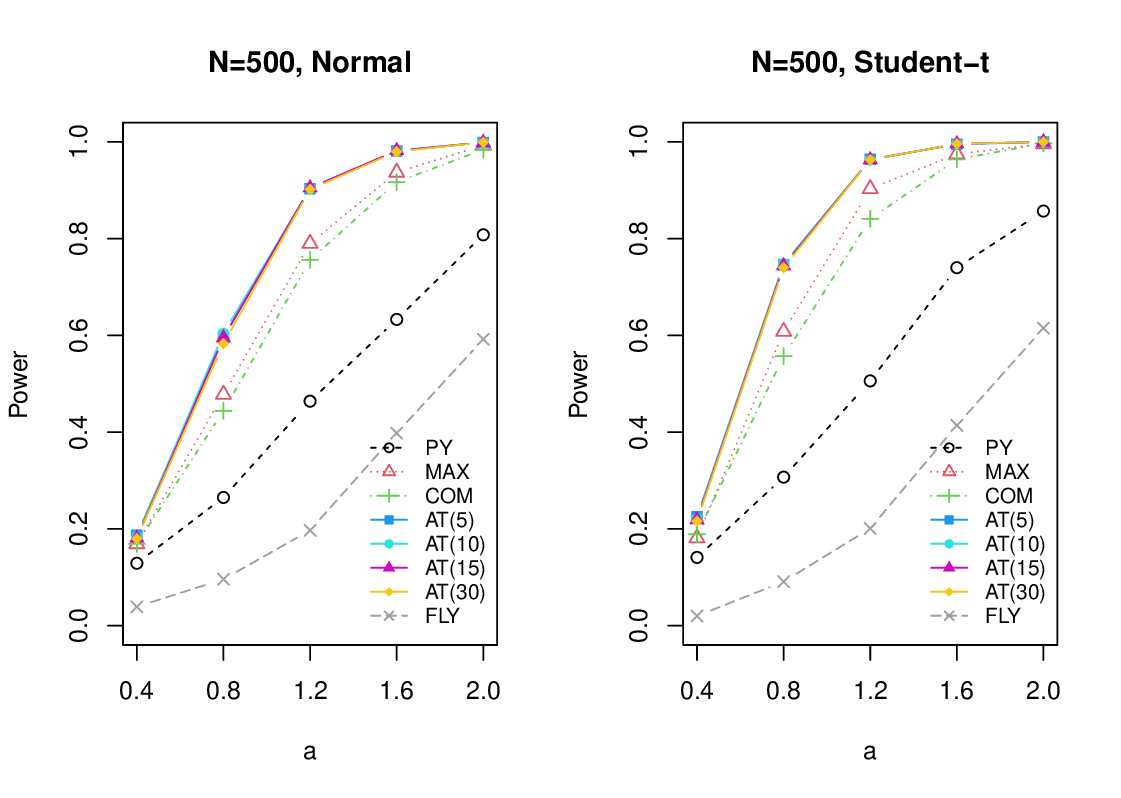}
	\caption{Power (\%) of different tests for $\Sigma_U$ in Case~(2) with $\delta_\gamma = 1/2$ and $k = \lfloor N^{1/3}\rfloor$}  	\label{FIG:17}
\end{figure}

\begin{figure}
	\centering
	\includegraphics[height=6cm,width=15cm]{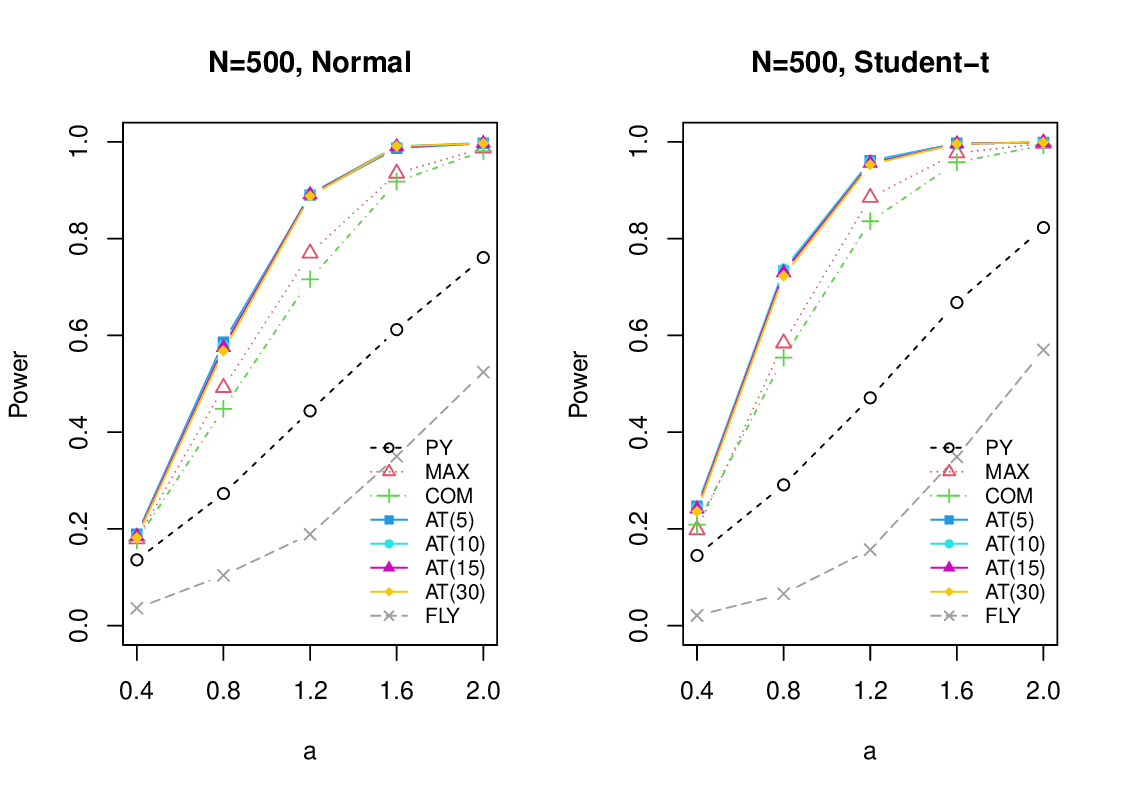}
	\caption{Power (\%) of different tests for $\Sigma_U$ in Case~(2) with $\delta_\gamma = 3/5$ and $k = \lfloor N^{1/3}\rfloor$}  	\label{FIG:18}
\end{figure}

We evaluate the performance of different procedures under the alternative hypothesis for $\Sigma_U$ in Case~(2) when $k = \lfloor N^{1/3}\rfloor$. The results are summarized in Figures~\ref{FIG:10}-\ref{FIG:18}. We observe that the proposed test consistently outperforms the other tests, indicating its robustness to strong cross-sectional dependence. We also find that the power of the PY and FLY tests decreases as $\delta_\gamma$ gets larger, which is a similar observation to the case of $\Sigma_U$ in Case~(1). This trend is further highlighted in Figures~10-18 in Supplementary material.
Additional simulation results are provided in the Supplementary material, where we consider the case where $\varepsilon_{t}$ follows an ARCH process. Table~1 and Figures 19-24 present the corresponding empirical rejection rates and power curves of the different tests.

%{\color{red}When $\varepsilon_{t}$ consists of ARCH processes, the similar results of the simulations are recorded
%in Supplementary material, see Figures~19-24 for details.}

It is important to highlight that the power of the PY and FLY tests is sensitive to the strength of the cross-sectional dependence, which is controlled by the parameter $\delta_\gamma$. Additionally, we found that the proposed test is relatively robust to the choice of $K$. In summary, our test provides reasonable control of the type I error and delivers better power than the other tests under weak signals.

\subsection{Real data analysis}

We examine the monthly excess returns on all the constituents of the S\&P 500 from the CRSP database from January 1980 to December 2012, as analyzed by \cite{Fan2015}. One of the key empirical findings reported in \cite{Fan2015} is that market inefficiency is predominantly caused by a small fraction of stocks with positive intercepts, rather than a large proportion of slightly mispriced assets. This finding provides empirical evidence of sparse alternatives.
Our goal here is to test market efficiency on a rolling window basis. For each month from December 1987 to December 2012, we calculate test statistics using the preceding 96 months' returns ($T=96$). A total of 400 stocks were considered for this study, and we only include stocks without missing observations in the past eight years in the data for each testing month. We fit the Fama and French three-factor model to the data in each rolling window ($l=12/1987,12/1988,\ldots, 12/2012$):
\begin{align}\label{e1}
r^{(l)}_{it} - f^{(l)}_{t}  = \alpha^{(l)}_i + \beta^{(l)}_{i,\text{MKT}}(\text{MKT}^{(l)}_t
- f^{(l)}_t )+ \beta^{(l)}_{i,\text{SMB}}\text{SMB}^{(l)}_t
+ \beta^{(l)}_{i,\text{HML}}\text{HML}^{(l)}_t + u_{it}^{(l)}
\end{align}
for $i = 1,\ldots, N$ and $t = l - 95,\dots, l$,
where $r_{it}$ is the return for stock $i$ at month $t$, $f_{t}$ is the risk free rate, and MKT, SMB and HML represent the market, size and value factors, respectively.

We apply the proposed approach and the other four completing procedures to examine the market efficiency hypothesis for the data in each rolling window. The screening set suggested by \cite{Fan2015} is computed, which shows that only a few significant nonzero ``alpha" components exist in seven rolling windows, which directly implies the presence of sparse or weak signals.
For instance, with $l=12/1992$, we select $284$ stocks (Window 1) without missing observations and fit the Fama-French three-factor model, and record the resulting $p$-values from all the tests in Table~\ref{tab2}. At the 5\% significant level, only the proposed test shows significant evidence against the null, while the other tests fail to reject it, indicating that $\boldsymbol{\alpha}$ is possibly sparse in this window. Similarly, for $l=12/1996$ (Window 2) based on $279$ stocks, the proposed adaptive test provides the strongest evidence against the null, followed by the MAX test. For $l=12/1999$ (Window 3) based on $241$ stocks, both the PY and the proposed adaptive tests reject the null hypothesis, while the other tests fail to reject it. Lastly, with $l=12/2011$ (Window 4) based on $170$ stocks, all the tests find significant evidence against the null.

To better understand these results and conduct a fair comparison among different procedures, we performed a simulation study. We generated data from the fitted factor models for each of the rolling windows. Specifically, for the $l$th rolling window, we generated data from either the null or alternative hypothesis.
 \begin{itemize}
\item [\rm S1.]
Let $(\hat{\alpha}^{(l)}_i, \hat{\beta}^{(l)}_{i,\text{MKT}}, \hat{\beta}^{(l)}_{i,\text{SMB}}, \hat{\beta}^{(l)}_{i,\text{HML}})$ be the estimates of the coefficients and $\hat{u}^{(l)}_{it}$ be the corresponding residuals. Under the null, we generate the excess returns as follows
\begin{align*}
r^{\ast}_{it} =  \hat{\beta}^{(l)}_{i,\text{MKT}}(\text{MKT}^{(l)}_t
- f^{(l)}_t )+ \hat{\beta}^{(l)}_{i,\text{SMB}}\text{SMB}^{(l)}_t
+ \hat{\beta}^{(l)}_{i,\text{HML}}\text{HML}^{(l)}_t+ \epsilon_t\hat{u}^{(l)}_{it},
\end{align*}
for $i = 1,2,\ldots,N$ and $t = l-95,\ldots,l$, where $\epsilon_t$ follows
the Rademacher distribution.
%two point distribution attaching masses
%$(\sqrt{5}+1)/2\sqrt{5}$ and $(\sqrt{5}-1)/2\sqrt{5}$ to the points $(-\sqrt{5}+1)/2$ and $(\sqrt{5}+1)/2$
%[\cite{Mammen1993}].
\item [\rm S2.] We let
$$\tilde{\alpha}^{(l)}_i=\hat{\alpha}^{(l)}_iI\{|\hat{\alpha}_i^{(l)}|>\hat{\sigma}_i\log(\log T)\sqrt{\log N}\}.$$
Under the alternative, we generate the data according to the following model
\begin{align*}
r^{\ast}_{it} = \tilde{\alpha}^{(l)}_i + \hat{\beta}^{(l)}_{i,\text{MKT}}(\text{MKT}^{(l)}_t
- f^{(l)}_t )+ \hat{\beta}^{(l)}_{i,\text{SMB}}\text{SMB}^{(l)}_t
+ \hat{\beta}^{(l)}_{i,\text{HML}}\text{HML}^{(l)}_t+ \epsilon_t\hat{u}^{(l)}_{it},
\end{align*}
for $i = 1, 2, \ldots, N$ and $t = l-95, \ldots, l$, where $\epsilon_t$ follows the Rademacher distribution.
\end{itemize}
We present the results of a simulation study in Table~\ref{tab2}, which allows us to compare the power and empirical size of the different tests. We conducted 1000 replications for each scenario and calculated the average power and empirical size. All tests showed reasonable control over the empirical size, although the FLY test was conservative. The proposed adaptive test was consistently the most powerful in all scenarios, outperforming the MAX and COM tests by approximately 10\% in scenarios D3 and D4. These findings are consistent with the results obtained from the real data analysis, indicating that the proposed test may be a useful addition to the existing methods for testing the market efficiency hypothesis.

\section*{Acknowledgments}

The authors thank the editor-in-chief, the associate editor, and the anonymous reviewers for their comments that significantly improved this work. The research of Qiang Xia was supported in part by the National Natural Science Foundation of China (No.12171161, 91746102) and the Natural
Science Foundation of Guangdong Province of China (No.2022A1515011754). The research of Xianyang Zhang was supported in part by NSF DMS-2210735.

\bigskip
\begin{center}
{\large\bf SUPPLEMENTARY MATERIAL}
\end{center}

\begin{description}

\item[Title:] The supplement contains some intermediate theoretical results and the proofs of the theorems.

\end{description}

\bibliographystyle{Chicago}

\bibliography{jbes-1}

\begin{table}[h]
 \caption{Size (\%) of different tests for $\Sigma_U$ in Case~(1) at the 5\% significance level}
 \label{tab1}
\begin{center}
\small
{\begin{tabular}{cccccccccccccccccc}
\hline
&&\multicolumn{3}{c}{$\delta_\gamma=1/4$}&&\multicolumn{3}{c}{$\delta_\gamma=1/2$}&&\multicolumn{3}{c}{$\delta_\gamma=3/5$}\\
\cline{3-5}\cline{7-9}\cline{11-13}
 &$N=$&$100$&$ 200$&$500$&&$100$&$200$&$500$&&$100$&$200$&$500$\\
\hline
Method &$T=100$&&&\multicolumn{4}{c}{Normal errors}\\
\hline
$\mathcal{T}_{\text{FLY}}$&    &5.1&5.9&5.3&&2.8&2.1&1.7&&0.6&0.4&0.3\\
%$\mathcal{T}_{\text{GL}}$&    &0.1&0.0&0.0&&0.2&0.0&0.0&&0.2&0.0&0.4\\
$\mathcal{T}_{\text{PY}}$&  &5.5&6.3&5.6&&5.4&4.4&5.4&&6.0&5.9&5.0\\
$\mathcal{T}_{\text{MAX}}$&    &4.9&5.6&7.1&&3.6&4.7&6.1&&3.9&4.4&6.4\\
$\mathcal{T}_{\text{COM}}$&    &4.6&5.5&6.3&&4.9&4.5&5.6&&5.8&5.9&6.0\\
$\widetilde{\mathcal{G}}_T(5)$&&7.2&7.5&6.5&&7.2&7.2&5.9&&6.4&6.6&6.2\\
$\widetilde{\mathcal{G}}_T(10)$&&6.9&7.0&6.5&&6.8&6.9&5.8&&6.0&6.1&6.3\\
$\widetilde{\mathcal{G}}_T(15)$&&6.7&6.7&6.4&&6.5&6.8&5.8&&5.8&6.0&6.2\\
$\widetilde{\mathcal{G}}_T(30)$&&6.7&6.6&6.3&&6.3&6.8&5.8&&5.6&5.9&6.1\\
\hline
 &$T=100$&&&\multicolumn{4}{c}{Student-t errors}\\
\hline
$\mathcal{T}_{\text{FLY}}$&    &4.0&3.6&2.9&&1.9&1.0&0.8&&0.2&0.2&0.3\\
%$\mathcal{T}_{\text{GL}}$&    &0.1&0.0&0.0&&0.2&0.0&0.0&&0.2&0.0&0.4\\
$\mathcal{T}_{\text{PY}}$&  &5.2&3.7&4.1&&5.4&5.0&3.9&&5.1&6.8&6.2\\
$\mathcal{T}_{\text{MAX}}$&    &3.0&3.8&3.3&&3.3&3.4&4.1&&2.4&3.0&4.4\\
$\mathcal{T}_{\text{COM}}$&    &3.8&4.2&3.5&&5.0&4.1&4.4&&4.2&5.6&6.2\\
$\widetilde{\mathcal{G}}_T(5)$&&5.3&5.0&3.1&&5.2&5.1&3.8&&4.1&3.7&3.8\\
$\widetilde{\mathcal{G}}_T(10)$&&4.6&4.7&3.1&&5.0&4.9&3.8&&4.0&3.8&3.8\\
$\widetilde{\mathcal{G}}_T(15)$&&4.4&4.5&3.1&&4.8&4.9&3.8&&4.0&3.8&3.8\\
$\widetilde{\mathcal{G}}_T(30)$&&4.4&4.2&3.1&&4.5&4.7&3.7&&4.1&3.7&3.8\\
\hline
\end{tabular}}
\end{center}
\end{table}

\clearpage

\clearpage
\clearpage
\clearpage
\clearpage
\clearpage
\clearpage

\begin{table}[h]
 \caption{Testing results of different procedures for the real data and simulated data mimicking the real data structure}
 \label{tab2}
\begin{center}
\small
{\begin{tabular}{cccccccccccc}
\hline
&Data $\backslash$ Test&$\mathcal{T}_{\text{PY}}$&$\mathcal{T}_{\text{MAX}}$&$\mathcal{T}_{\text{COM}}$&
$\widetilde{\mathcal{G}}_T(5)$ & $\widetilde{\mathcal{G}}_T(10)$ &$\widetilde{\mathcal{G}}_T(30)$
&$\mathcal{T}_{\text{FLY}}$\\
\hline
\hline
&Window~1&&&& $p$-value &&&\\
\hline
&&0.0942&0.0721&0.0832&0.0240&0.0240&0.0260&0.1267\\
\hline
$H_0$&Bootstrap data&&&& Size (\%) &&&\\
\hline
& &8.6&4.5&7.8&5.8&5.9&5.7&2.6\\
\hline
$H_a$&Bootstrap data&&&& Power (\%) &&&\\
\hline
& &18.2&69.6&65.0&74.4&74.5&74.4&9.2\\
\hline
\hline
&Window~2&&&& $p$-value &&&\\
\hline
&&0.1189&0.0432&0.0861&0.0140&0.0140&0.0140&0.1368\\
\hline
$H_0$&Bootstrap data&&&& Size (\%) &&&\\
\hline
& &6.4&5.3&6.4&5.6&5.7&5.6&1.6\\
\hline
$H_a$&Bootstrap data&&&& Power (\%) &&&\\
\hline
& &12.3&59.8&53.4&63.3&63.2&63.1&4.4\\
\hline
\hline
&Window~3&&&& $p$-value &&&\\
\hline
&&0.0364&0.0936&0.0650&0.0160&0.0160&0.0180&0.1005\\
\hline
$H_0$&Bootstrap data&&&& Size (\%) &&&\\
\hline
& &8.7&5.3&8.2&5.5&5.5&5.4&2.0\\
\hline
$H_a$&Bootstrap data&&&& Power (\%) &&&\\
\hline
& &14.2&55.3&49.2&60.8&60.8&60.6&3.2\\
\hline
\hline
&Window~4&&&& $p$-value &&&\\
\hline
&&0.0035&0.0168&0.0102&0.0000&0.0000&0.0000&0.0482\\
\hline
$H_0$&Bootstrap data&&&& Size (\%) &&&\\
\hline
& &8.1&4.2&6.8&4.4&4.4&4.4&2.6\\
\hline
$H_a$&Bootstrap data&&&& Power (\%) &&&\\
\hline
& &21.4&73.3&68.1&79.8&79.6&79.6&10.2\\
\hline
\end{tabular}}
\end{center}
\end{table}

\end{document}